\documentclass[acmsmall,screen]{acmart}

\def\BibTeX{{\rm B\kern-.05em{\sc i\kern-.025em b}\kern-.08emT\kern-.1667em\lower.7ex\hbox{E}\kern-.125emX}}

\usepackage{caption}
\usepackage{subcaption}
\usepackage{hyperref}
\usepackage{listings}
\usepackage{xcolor}
\usepackage[utf8]{inputenc}
\usepackage{lipsum} 
\usepackage{xifthen}
\usepackage{pifont} 
\usepackage{array}  

\definecolor{codegreen}{rgb}{0,0.6,0}
\definecolor{codegray}{rgb}{0.5,0.5,0.5}
\definecolor{codepurple}{rgb}{0.58,0,0.82}
\definecolor{backcolour}{rgb}{0.98,0.98,0.98}
\definecolor{bg}{HTML}{F8F9FB}  

\lstdefinestyle{mystyle}{
    backgroundcolor=\color{backcolour},   
    commentstyle=\color{codegreen},
    keywordstyle=\color{magenta},
    numberstyle=\tiny\color{codegray},
    stringstyle=\color{teal},
    basicstyle=\ttfamily\footnotesize\linespread{0.8},
    breakatwhitespace=false,         
    breaklines=true,                 
    captionpos=b,                    
    keepspaces=true,                 
    numbers=left,                    
    numbersep=5pt,                  
    showspaces=false,                
    showstringspaces=false,
    showtabs=false,                  
    tabsize=2,
    xleftmargin=8pt,
    lineskip={-2.0pt}
}

\usepackage[framemethod=TikZ]{mdframed}
\mdfdefinestyle{MyFrame}{%
    linecolor=black,
    outerlinewidth=0.3pt,
    roundcorner=5pt,
    skipabove = 5.5pt,
    skipbelow = 5.5pt,
    innertopmargin=5.5pt, 
    innerbottommargin=5.5pt, 
    innerrightmargin=5.5pt,
    innerleftmargin=5.5pt,
    backgroundcolor=bg, 
}

\definecolor{bgGreen}{RGB}{39, 110, 64}
\definecolor{bgRed}{RGB}{214, 64, 64}

\usepackage{enumitem}
\usepackage{multirow}
\usepackage{xspace}
\usepackage[ruled,lined,linesnumbered,vlined]{algorithm2e}
\usepackage{graphicx}
\usepackage{color}
\usepackage{hyperref}
\usepackage{balance}
\usepackage{siunitx}
\usepackage{adjustbox}
\usepackage{booktabs}
\usepackage{bchart}
\usepackage[framemethod=TikZ]{mdframed}
\usepackage{color,colortbl}
\usepackage{comment}
\definecolor{Gray}{gray}{0.9}
\definecolor{darkgreen}{rgb}{0,0.5,0}
\usepackage{tikz}
\usepackage{pgfplots}
\usepackage[frozencache,cachedir=minted-cache]{minted}
\usepackage{booktabs,caption}

\usepackage[para,online,flushleft]{threeparttable}
\usepackage{etoolbox}
\appto\TPTnoteSettings{\footnotesize}

\usepackage{calc}
\newcommand*{\mycode}{\fontfamily{lmtt}\selectfont}

\newcommand{\eg}{\hbox{\emph{e.g.}}\xspace}
\newcommand{\ie}{\hbox{\emph{i.e.}}\xspace}

\definecolor{mygreen}{HTML}{02818a}
\definecolor{bg}{HTML}{F8F9FB}  
\definecolor{bgc}{HTML}{FCF6E4}

\newcommand*{\alreadyfixedbugsvalue}{10}
\newcommand*{\alreadyconfirmedbugsvalue}{3}
\newcommand*{\knownbugsvalue}{\the\numexpr\alreadyconfirmedbugsvalue+\alreadyfixedbugsvalue}
\newcommand*{\pendingbugsvalue}{0}
\newcommand*{\rejectedbugsvalue}{12}
\newcommand*{\unconfirmedbugsvalue}{\the\numexpr\pendingbugsvalue+\rejectedbugsvalue}
\newcommand*{\prbugsvalue}{13}
\newcommand*{\noprbugsvalue}{23}
\newcommand*{\newnotfixedbugsvalue}{\the\numexpr\prbugsvalue+\noprbugsvalue}
\newcommand*{\fixedbugsvalue}{10}
\newcommand*{\newbugsvalue}{\the\numexpr\newnotfixedbugsvalue+\fixedbugsvalue}
\newcommand*{\confirmedbugsvalue}{\the\numexpr\knownbugsvalue+\newbugsvalue}
\newcommand*{\reportedbugsvalue}{\the\numexpr\unconfirmedbugsvalue+\confirmedbugsvalue}

\newcommand*{\knownbugs}{\knownbugsvalue\xspace}

\newcommand*{\reportedbugs}{\reportedbugsvalue\xspace}
\newcommand*{\fixedbugs}{\fixedbugsvalue\xspace}

\newcommand*{\rejectedbugs}{\rejectedbugsvalue\xspace}

\newcommand*{\confirmedbugs}{\confirmedbugsvalue\xspace}
\newcommand*{\newnotfixedbugs}{\the\numexpr\newnotfixedbugsvalue\xspace}
\newcommand*{\newbugs}{\the\numexpr\newbugsvalue\relax\xspace}

\newcommand{\myparagraph}[1]{
  \noindent \textit{\textbf{#1.}}
}

\usepackage{amsmath}
\usepackage{algpseudocode}

\newcommand*{\toolname}{\textsc{DLLens}\xspace}
\newcommand*{\toolnamenonllm}{\textsc{DLLens} (w/o ICF)\xspace}
\newcommand*{\llm}{{\mycode{}gpt-4o-mini}\xspace}

\newcommand*{\undertest}{$f$\xspace}
\newcommand*{\counterpart}{$f'$\xspace}
\newcommand*{\validinputs}{$\mathcal{X}$\xspace}
\newcommand*{\mutants}{$\mathcal{X}_{mutants}$\xspace}
\newcommand*{\validmutants}{$\mathcal{X}_{validmutants}$\xspace}
\newcommand{\mypath}{\ensuremath{p}\xspace}
\newcommand{\myinput}{\ensuremath{I}\xspace}

\newcommand{\myPC}{\ensuremath{PC}\xspace}
\newcommand{\crash}{\textit{crash}\xspace}
\newcommand{\inconsistent}{\textit{inconsistent}\xspace}

\usepackage{cleveref} 

\SetKwData{Stmt}{stmt}
\SetKwData{SanityCheckStmt}{SanityCheckStmt}
\SetKwData{ConditionalStmt}{ConditionalStmt}
\SetKwData{LocalCons}{$condition$}
\SetKwData{InputCons}{$Cons_{input}$}
\SetKwData{ErrorCons}{$Cons_{error}$}
\SetKwData{NaturalCons}{$Cons_{natural}$}
\SetKwData{PropertyCons}{$Cons_{property}$}
\SetKwData{DuplicateCons}{$Cons_{duplicate}$}
\SetKwData{PCPool}{$Pool_{\myPC}$}
\SetKwData{DF}{$Flow_{\myinput\rightarrow\LocalVars}$}
\SetKwData{LocalVars}{$Var_{local}$}
\SetKwData{InputVars}{$Var_{I}$}
\SetKwData{PCSelector}{$Selector_{PC}$}
\SetKwData{TestInput}{$\myinput_{test}$}
\SetKwData{FuzzingDriver}{$driver$}
\SetKwFunction{AlgPCExtraction}{PathConstraintExtraction}
\SetKwFunction{AlgPCSelection}{PathConstraintSelection}
\SetKwFunction{AlgTestInputGeneration}{TestInputGeneration}
\SetKwFunction{AlgConstraintConstruction}{Cons.Construct}
\SetKwFunction{AlgTraverse}{Traverse}
\SetKwFunction{AlgInputConstraintInference}{Inp.Cons.Inf.}
\SetKwFunction{AlgExtractConstraint}{ConstructConstraint}
\SetKwFunction{AlgSelfValid}{Self-Validation}
\SetKwFunction{AlgSyntaxValid}{IsSyntaxValid}
\SetKwFunction{AlgContainExternal}{HasExternal}
\SetKwFunction{AlgSolvable}{Solvable}

\Crefname{algocf}{Algorithm}{Algorithms}
\crefname{algocf}{Algorithm}{Algorithms}

\Crefname{algorithm}{Algorithm}{Algorithms}
\crefname{algorithm}{Algorithm}{Algorithms}

\crefname{appendix}{Appendix}{Appendices}
\Crefname{appendix}{Appendix}{Appendices}

\Crefname{figure}{Figure}{Figures}
\crefname{figure}{Figure}{Figures}

\crefname{listing}{Listing}{Listings}
\Crefname{listing}{Listing}{Listings}

\Crefname{table}{Table}{Tables}
\crefname{table}{Table}{Tables}

\crefname{thm}{Theorem}{Theorems}
\Crefname{thm}{Theorem}{Theorems}

\crefformat{chapter}{\S#2#1#3}
\crefmultiformat{chapter}{\S\S#2#1#3}{ and~#2#1#3}{, #2#1#3}{, and~#2#1#3}

\crefformat{section}{\S#2#1#3}
\crefmultiformat{section}{\S\S#2#1#3}{ and~#2#1#3}{, #2#1#3}{, and~#2#1#3}

\begin{document}

\title{Enhancing Differential Testing With LLMs For Testing Deep Learning Libraries}

\author{Meiziniu Li}
\orcid{0000-0001-5947-4030}  
\affiliation{
  \institution{The Hong Kong University of Science and Technology}
  \country{Hong Kong, China}
}
\email{mlick@cse.ust.hk}

\author{Dongze Li}
\orcid{0009-0004-6482-8509}  
\affiliation{
  \institution{The Hong Kong University of Science and Technology}
  \country{Hong Kong, China}
}
\email{dlibk@connect.ust.hk}

\author{Jianmeng Liu}
\orcid{0009-0004-8423-7781}  
\affiliation{
  \institution{The Hong Kong University of Science and Technology}
  \country{Hong Kong, China}
}
\email{jliudq@connect.ust.hk}

\author{Jialun Cao}
\orcid{0000-0003-4892-6294}  
\affiliation{
  \institution{The Hong Kong University of Science and Technology}
  \country{Hong Kong, China}
}
\email{jcaoap@cse.ust.hk}

\author{Yongqiang Tian}
\orcid{0000-0003-1644-2965}  
\affiliation{
  \institution{Monash University}
  \country{Melbourne, Australia}
}
\email{yongqiang.tian@monash.edu}

\author{Shing-Chi Cheung*}
\orcid{0000-0002-3508-7172}  
\thanks{* Corresponding author.}
\affiliation{
  \institution{The Hong Kong University of Science and Technology}
  \country{Hong Kong, China}
}
\email{scc@cse.ust.hk}

\begin{abstract}
Differential testing offers a promising strategy to alleviate the test oracle problem by comparing the test results between alternative implementations. However, existing differential testing techniques for deep learning (DL) libraries are limited by the key challenges of finding alternative implementations (called $counterparts$) for a given API and subsequently generating diverse test inputs.
To address the two challenges, this paper introduces \toolname, an LLM-enhanced differential testing technique for DL libraries.  
The first challenge is addressed by an observation that DL libraries are commonly designed to support the computation of a similar set of DL algorithms. 
Therefore, the counterpart of a given API's computation could be successfully synthesized through certain composition and adaptation of the APIs from another DL library. 
\toolname{} incorporates a novel counterpart synthesis workflow, leveraging a large language model (LLM) to search for valid counterparts for differential testing.
To address the second challenge, \toolname{} incorporates a static analysis technique that extracts the path constraints from the implementations of a given API and its counterpart to guide diverse test input generation. 
The extraction is facilitated by LLM's knowledge of the concerned DL library and its upstream libraries. 
\toolname{} incorporates validation mechanisms to manage the LLM's hallucinations in counterpart synthesis and path constraint extraction.

We evaluate \toolname{} on two popular DL libraries, TensorFlow and PyTorch.
Our evaluation shows that \toolname{} synthesizes counterparts for 1.84 times as many APIs as those found by state-of-the-art techniques on these libraries.
Moreover, under the same time budget, \toolname{} covers 7.23\% more branches and detects 1.88 times as many bugs as state-of-the-art techniques on 200 randomly sampled APIs.
\toolname{} has successfully detected \reportedbugs{} bugs in recent TensorFlow and PyTorch libraries. 
Among them, \confirmedbugs{} are confirmed by developers, including \newbugs{} confirmed as previously unknown bugs, and \fixedbugs{} of these previously unknown bugs have been fixed in the latest version of TensorFlow and PyTorch.
\end{abstract}



\maketitle

\section{Introduction}\label{sec:introduction}
Deep learning (DL) has been increasingly adopted for mission-critical applications such as authentication~\cite{das2018deep,Ferdowsi2019DeepLF}, medical treatment~\cite{litjens2017survey}, and autonomous driving~\cite{autonomousdriving,Sallab2017DeepRL,ShalevShwartz2016SafeMR}.
Modern DL applications are mostly developed on top of popular DL libraries such as PyTorch~\cite{pytorch} and TensorFlow~\cite{tensorflow}.
However, the presence of functional incorrectness in DL libraries, commonly referred to as functional bugs, poses a significant threat to the reliability of DL applications~\cite{silentbugs,empirical_2022_dlframeworks,zhang2018empirical}. 
More than 30\% of bugs reported to PyTorch developers are categorized as functional bugs (\eg, incorrect computation results or execution states)~\cite{chen2022toward}.
Hence, functional bug detection is critical to ensure the quality of DL libraries.

Differential testing is commonly adopted~\cite{cradle,lemon,deeprel,muffin,comet,tensorscope} to detect functional bugs.
A general workflow of these works is to (1) identify the counterparts (\ie, different implementations that offer the same functionality) in/across DL libraries, and (2) detect output discrepancy between counterparts via test inputs.
However, existing works are limited in both steps, which may hinder their effectiveness in bug detection. 

\myparagraph{(1) Limitation of finding counterparts for differential testing}
Several approaches have been proposed to find counterparts of DL library APIs for differential testing.
The counterparts are either another similar API (\eg, {\mycode{}tf.math.argmax} and {\mycode{}tf.argmax}) in the same DL library~\cite{deeprel} or the API's different computational modes, such as \textit{hardware backends} (\eg, CPU and GPU)~\cite{freelunch,titanfuzz,fuzzgpt} and \textit{execution modes} (\eg, different gradient calculation modes)~\cite{lambdafuzz, eagle}.
However, these counterparts often have shared underlying implementations~\cite{deeprel,tensorscope},
thus may yield identical computation results, making differential testing ineffective~\cite{tensorscope}.
To address this problem, a recent approach, TensorScope~\cite{tensorscope}, finds API's counterparts from different DL libraries.
TensorScope extracts counterparts from existing sources (\ie, counterparts identified by existing model conversion libraries such as TF2ONNX).
Although this strategy is less likely to collect counterparts producing identical computation results, its effectiveness is limited by the small set of APIs supported by these model conversion libraries~\cite{freelunch} (see \cref{subsec:limitation_counterparts}).
Specifically, TensorScope can only collect counterparts for 9.4\% (304/3,249) TensorFlow APIs~\cite{tensorscope}, leaving 90\%+ of TensorFlow APIs not covered.
Listing~\ref{lst:motivating_bug} shows a real functional bug~\cite{tf_is_non_decreasing} that TensorScope and other approaches cannot trigger. 
Existing approaches are ineffective in detecting this bug since none of the conversion libraries support this API ({\mycode{}tf.math.is\_non\_decreasing}), and the counterparts extracted from the same DL library consistently deliver incorrect outputs (see \cref{subsec:limitation_counterparts}).

\myparagraph{(2) Limitation of generating test inputs to explore diverse execution paths}
Existing DL library testing approaches commonly rely on input constraints for test input generation.
These approaches either define input constraints manually~\cite{predoo,comet} or extract them from various sources, including API documentation~\cite{docter} and input assertions~\cite{tensorscope}.
However, these input constraints mainly focus on ensuring input validity rather than guiding test input generation to explore diverse execution paths.
To address this problem, ACETest~\cite{acetest} proposes using static analysis to extract path constraints, each of which is a set of input constraints required by a specific execution path in the source code of DL library APIs. 
These path constraints facilitate test input generation for diverse path exploration.
Yet, due to the large codebase of DL library, ACETest does not perform static analysis on some external functions (\ie, functions from external libraries or modules) used by DL library APIs, which may result in some incomplete path constraints~\cite{acetest}.
Listing~\ref{lst:explore_path} demonstrates the limitation of existing approaches, where line 15, performs the core logic of the API {\mycode{}tf.raw\_ops.MatrixDiagV2}, is difficult to be reached by these approaches.
To reach this line, a test input must pass three sanity checks (lines 3, 13--14) and one {\mycode{}If} statement (line 4).
However, only part of the constraints required for reaching line 15 can be captured by existing approaches (see \cref{subsec:limitation_constraint} for details).
Many bugs in DL library APIs reside deeply at the core logic, which requires test executions to traverse various branches and satisfy multiple sanity checks~\cite{acetest}.
Extracting more comprehensive path constraints can facilitate the test input generation toward exploring diverse execution paths, thereby enhancing the chances of detecting bugs.

\begin{listing}
\begin{minted}[
    baselinestretch=1.0,
    fontsize=\scriptsize,
    xleftmargin=0.5ex,
    bgcolor=bg,
    breaklines=true,
    escapeinside=||,
]{text}
|\underline{\textbf{Bug Triggering Input}}|:
x = tf.constant([10,9], dtype='uint32')
|\underline{\textbf{Buggy API}}|:
tf.math.is_non_decreasing
|\underline{\textbf{Actual Result (Expected Result)}}|:
|\textbf{\textcolor{bgRed}{True}}| (|\textbf{\textcolor{bgGreen}{False}}|)
|\underline{\textbf{Developer's Fix}}|:
|\textbf{\textcolor{bgRed}{- diff = \_get\_diff\_for\_monotonic\_comparison(x)}}|
|\textbf{\textcolor{bgRed}{- zero = ops.convert\_to\_tensor(0, dtype=diff.dtype)}}|
|\textbf{\textcolor{bgRed}{- return math\_ops.reduce\_all(math\_ops.less\_equal(zero, diff))}}|
|\textbf{\textcolor{bgGreen}{+ diff = \_get\_results\_for\_monotonic\_comparison(x, greater\_equal)}}|
|\textbf{\textcolor{bgGreen}{+ return math\_ops.reduce\_all(diff)}}|
\end{minted}
\caption{A real bug detected by \toolname{} that leads to an incorrect computation result. 
}
\label{lst:motivating_bug}
\end{listing}

\begin{listing}
\begin{minted}[
    baselinestretch=1.0,
    fontsize=\scriptsize,
    xleftmargin=0.5ex,
    bgcolor=bg,
    breaklines=true,
    escapeinside=??,
]{c++}
?[\textbf{Input}]?
diagonal, diag_index, num_rows_tensor, num_cols_tensor, ...
?[\textbf{Code Snippet (Truncated)}]?
1:  lower_diag_index = diag_index.flat<int32>()(0);
2:  upper_diag_index = lower_diag_index;
3:  OP_REQUIRES(context, diag_index.NumElements() > 0,...); // ------------------- ?\textbf{Sanity Check 1 (SC1)}?
4:  if (diag_index.dim_size(0) > 1) {  // ---------------------------------------- ?\textbf{If Statement}?
5:      upper_diag_index = diag_index.flat<int32>()(1);    // -------------------- ?\textbf{True Branch 1 (TB1)}?
    }                                                      // -------------------- ?\textbf{False Branch 1 (FB1)}?
6:  num_rows = num_rows_tensor.flat<int32>()(0);
7:  num_cols = num_cols_tensor.flat<int32>()(0);
8:  const TensorShape& diagonal_shape = diagonal.shape();
9:  const int diag_rank = diagonal_shape.dims();
10: const Eigen::Index max_diag_len = diagonal_shape.dim_size(diag_rank - 1);
11: const int32_t min_num_rows = max_diag_len - ?\textbf{\underline{std::min}}?(upper_diag_index, 0);  // ?\underline{\textbf{External Function}}?
12: const int32_t min_num_cols = max_diag_len + ?\textbf{\underline{std::max}}?(lower_diag_index, 0);  // ?\underline{\textbf{External Function}}?
13: OP_REQUIRES(context, num_rows == -1 || num_rows >= min_num_rows, // ---------- ?\textbf{Sanity Check 2 (SC2)}?
            errors::InvalidArgument("The number of rows is too small."));
14: OP_REQUIRES(context, num_cols == -1 || num_cols >= min_num_cols, // ---------- ?\textbf{Sanity Check 3 (SC3)}?
            errors::InvalidArgument("The number of columns is too small."));
    ...
15: functor::MatrixDiag<Device, T>::Compute(...);  // ---------------------------- ?\textbf{Core logic}?
\end{minted}
\caption{
    A modified code snippet in API {\mycode{}tf.raw\_ops.MatrixDiagV2}.
    Test inputs need to pass three sanity checks (lines 3, 13--14) and one {\mycode{}If} statement (line 4) before reaching the core logic at line 15.
    \underline{\textbf{Underlined functions}} refer to external functions.
    }
\label{lst:explore_path}
\end{listing}

\myparagraph{Solution}
This paper introduces \toolname, an LLM-enhanced differential testing technique for DL library API testing.
\toolname{} is designed to effectively find counterparts and generate test inputs to explore diverse execution paths with the help of an LLM\@.
The insight of finding counterparts is that APIs in DL libraries are commonly designed to support various computations for a similar set of published DL algorithms (\eg, the convolution operation).
While these APIs may not have one-to-one mappings, their computations can be simulated mutually through certain API compositions and adaptations. 
Leveraging this insight, \toolname{} utilizes an LLM to synthesize counterparts via composing and adapting APIs from a different DL library\@.
To mitigate the hallucination problem of the LLM, \toolname{} incorporates a novel counterpart synthesis workflow that validates the synthesized counterparts and guides the LLM toward synthesizing a valid counterpart for each API under test.
Taking the buggy API in Listing~\ref{lst:motivating_bug} as an example, despite PyTorch lacking an API equivalent to this buggy API, \toolname{} successfully synthesizes a valid counterpart of this API using a combination of three PyTorch APIs (see Listing~\ref{lst:motivating_counterpart}).
This synthesized counterpart further facilitates \toolname{} to detect the incorrect computation result of the buggy API in Listing~\ref{lst:motivating_bug}.
To guide test input generation toward exploring diverse execution paths, \toolname{} proposes an LLM-enhanced static analysis method to extract path constraints for execution paths\@.
One challenge of extracting path constraints is that static analysis can become prohibitively expensive when applied to the large codebase of external functions and their dependencies.
Both skipping these external functions or manually modeling a limited set of them may result in incomplete path constraints.
\toolname{} addresses this challenge by employing an LLM to infer constraints involving these external functions. 
This is motivated by an insight that LLMs have learned domain-specific knowledge about DL libraries and their upstream libraries. 
Thus, these LLMs may likely have partial knowledge about the constraints imposed by these external functions.
A validation mechanism is subsequently applied to validate the constraints inferred by the LLM. The mechanism aims to avoid using invalid constraints inferred by the LLM for test input generation.
Lastly, \toolname{} applies a test input generation method using path constraints extracted from the implementation of each DL library API and its counterpart.
Differential testing is further applied by comparing outputs between the API and its counterpart.

\begin{listing}
\begin{minted}[
    baselinestretch=1.0,
    fontsize=\scriptsize,
    xleftmargin=0.5ex,
    bgcolor=bg,
    breaklines=true,
    escapeinside=||,
]{text}
|\underline{\textbf{TensorFlow API}}|
tf.math.is_non_decreasing(x)
|\underline{\textbf{Function Using PyTorch's APIs}}|
def pytorch_call(x):
    return torch.all(torch.eq(x, torch.sort(x)[0]))
\end{minted}
\caption{A function composing PyTorch's APIs can implement the specified computation of the API \\{\mycode{}tf.math.is\_non\_decreasing}.}
\label{lst:motivating_counterpart}
\end{listing}

We evaluate the effectiveness of \toolname{} on two popular DL libraries: TensorFlow~\cite{tensorflow} and PyTorch~\cite{pytorch}.
We compare \toolname{} with existing DL library testing techniques on the effectiveness of counterpart synthesis, path constraint extraction, branch coverage, and bug detection.
Our evaluation shows that \toolname{} finds 1.84 times (1401 v.s. 762) as many API counterparts as those found by the state-of-the-art technique~\cite{tensorscope}, which relies on counterparts identified by model conversion libraries.
Moreover, we notice that path constraints extracted through LLM-enhanced static analysis can better guide test input generation to reach more library branches, compared with our static analysis without using LLM\@.
Our experiment on 200 randomly sampled APIs further shows that \toolname{} can cover at least 7.23\% more branches and detect at least 1.88 times as many bugs as the state-of-the-art techniques.

In total, \toolname{} detected \reportedbugs{} bugs, including \newbugs{} confirmed as previously unknown ones after we reported them.
So far, \fixedbugs{} of our detected previously unknown bugs have been fixed in the latest version of TensorFlow and PyTorch.

This work makes the following contributions{:} 
\begin{itemize}[leftmargin=*,topsep=0pt]
    \item It proposes an automated technique, \toolname, to deploy differential testing with the help of LLMs for DL library fault detection.
    \toolname{} is the first technique that proposes to synthesize counterparts for DL library APIs by composing and adapting APIs from different DL libraries using LLMs.
    Additionally, \toolname{} incorporates an effective mechanism leveraging LLMs to address external functions encountered during static analysis, facilitating path constraint extraction in DL libraries.
    
    \item We implement \toolname{} and perform an extensive evaluation.
    Evaluation results indicate that \toolname{} finds 1.84 times (1401 v.s. 762) as many API counterparts as those found by the state-of-the-art technique, thereby applying differential testing to a wider range of DL library APIs. The results further demonstrate \toolname{}'s effectiveness in guiding test input generation, enabling the exploration of diverse execution paths.

    \item The application of \toolname{} to the recent releases of two popular DL libraries, TensorFlow and PyTorch, successfully detected \reportedbugs{} bugs. 
    Most (\newbugs) bugs have been confirmed as previously unknown bugs after we reported them. So far, \fixedbugs{} of our reported new bugs have been fixed in the latest version of TensorFlow and PyTorch.
    \item \textbf{Artifact Availability.} We make our artifact available online~\cite{DLLens}.
\end{itemize}

\section{Background and Motivation}\label{sec:moti}
\subsection{Differential Testing for DL Library API Testing}
To address the test oracle problem in DL library API testing, prior works~\cite{cradle,audee,deeprel,freelunch,titanfuzz,tensorscope} commonly employ differential testing for a target API \undertest{}.
These techniques identify potential bugs when the inconsistency between implementations exceeds a specified tolerance:
\begin{equation}
|f(x)-f'(x)| > \epsilon
\end{equation}
where $x$ (denoted as test input) represents concrete input parameters (including tensors) of \undertest{}, \counterpart{} (denoted as a counterpart) refers to a functionally equivalent implementation of \undertest{}, and $\epsilon$ is a user-defined threshold typically ranging between $10^{-1}$ and $10^{-3}$ across different techniques.

Research efforts to enable this paradigm have focused on two core challenges: (1) identifying counterparts for \undertest{}, and (2) extracting constraints for effective test input generation. 
While existing techniques have advanced the state of practice, limitations persist in both dimensions.
The rest of this section analyzes these limitations, followed by our analysis of the opportunities and challenges in leveraging LLMs to address them.

\subsection{Limitation of Existing DL Library API Counterpart Collection Approaches}\label{subsec:limitation_counterparts}
A counterpart of a DL library API \undertest{} is a function \counterpart{} simulating the intended computation of \undertest.
Recent approaches~\cite{freelunch,lambdafuzz,eagle,titanfuzz,fuzzgpt} consider different computational modes, such as different backends or different gradient calculation modes, as counterparts.
Besides these approaches, DeepREL~\cite{deeprel} considers APIs with similar names or similar signatures as counterparts.
Nevertheless, these counterparts often have shared implementations~\cite{tensorscope}, suggesting that they might yield identical outputs.
Taking API {\mycode{}tf.math.is\_non\_decreasing} in Listing~\ref{lst:motivating_bug} as an example, all counterparts collected by existing approaches (rows 1--3 in Table~\ref{tab:existing_counterparts}) also produced the same incorrect output value (\ie, {\mycode{}True}) for the bug triggering input. As a result, differential testing is ineffective in exposing the API implementation bug using the collected counterparts.

\begin{table}[t!]
\caption{Existing approaches' counterparts and their outputs on the bug triggering input for the buggy API {\mycode{}tf.math.is\_non\_decreasing(x)} in Listing~\ref{lst:motivating_bug}, along with result of our tool \toolname. The buggy API incorrectly outputs a \texttt{True} value for the input. 
FreeFuzz and $\nabla$Fuzz collect two different computation modes of the buggy API as counterparts.
}\label{tab:existing_counterparts}
\renewcommand{\arraystretch}{1.0}
\resizebox{0.7\linewidth}{!}{
\begin{tabular}{l|l|l}  
\toprule
\textbf{Approach}  &  \textbf{Counterparts Collected by the Approach} & \textbf{Output Value} \\ \midrule
FreeFuzz &  CPU / GPU &  \texttt{True} \\ \midrule
$\nabla$Fuzz &  REV / FWD  & \texttt{True}  \\ \midrule
DeepREL & {\mycode{}tf.compat.v1.is\_non\_decreasing(x)} &  \texttt{True}  \\ \midrule  
\toolname{} & {\mycode{}torch.all(torch.eq(x,torch.sort(x)[0]))} &  \texttt{False}  \\ \bottomrule  
\end{tabular}
}
\end{table}

To facilitate differential testing, we follow previous works~\cite{tensorscope,cradle,lemon,muffin,comet} and focus only on those counterparts with a different implementation from \undertest.
A practical way is to collect counterparts of \undertest{} from a different library (\ie, cross-library counterparts)~\cite{tensorscope}.
To do so, TensorScope~\cite{tensorscope} extracts counterparts identified by existing model conversion libraries by analyzing their conversion rules.
\textbf{However, when an API's counterpart is not included in model conversion libraries, no counterpart can be identified.}
Notably, we observe that current model conversion libraries only include counterparts for 304 out of 3,249 TensorFlow APIs and 458 out of 1,574 PyTorch APIs.
One explanation for this limited API coverage is that model conversion libraries primarily cater to layer APIs, such as {\mycode{}tf.keras.layers.Conv2D}, with less emphasis on other types like linear algebra or image processing APIs.
For instance, all model conversion libraries do not include a counterpart for {\mycode{}tf.math.is\_non\_decreasing}, which is an API for linear algebra.
Without a counterpart, differential testing cannot be performed to detect the output incorrectness of this API, thus is limited in detecting the bug in Listing~\ref{lst:motivating_bug}\@.

\subsection{Limitation of Existing DL Library Constraint Extraction Approaches}\label{subsec:limitation_constraint}
Previous approaches commonly extract input constraints from various sources to guide test input generation, including API documentation and source code.
DocTer~\cite{docter} employs rule-based approaches to extract input constraints from the API documentation.
TensorScope~\cite{tensorscope} analyzes APIs' source code and extracts constraints from sanity checks within these source code.
However, both approaches mainly focus on ensuring the generated input's validity rather than extracting input constraints required by specific execution paths. 
Thus, they are limited in generating test inputs to explore various execution paths.
Additionally, constraints generated by these approaches may lack comprehensiveness, as DocTer only relies on documentation, and TensorScope's analysis is restricted to a subset of sanity checks.

ACETest~\cite{acetest} is the state-of-the-art approach to extract path constraints in DL library APIs.
For each execution path, ACETest uses static analysis to extract input constraints required by branch conditions and sanity checks.
However, it only performs static analysis on a subset of API source code, excluding the source code of external functions from external libraries or modules.
\textbf{As external functions are prevalent in the implementation of DL library APIs, omitting them can result in incomplete path constraints.}
For instance, the code (lines 11--12) in Listing~\ref{lst:explore_path} includes external functions named {\mycode{}std::min} and {\mycode{}std::max}, which ACETest does not analyze.
Without analyzing these external functions, the input constraints required by the following sanity checks (lines 13--14) cannot be captured by ACETest.

\begin{table}[t!]
\caption{Constraints extracted by existing approaches for the code snippet in~\cref{lst:explore_path}, along with the result of our tool \toolname. Columns 3--7 indicate whether constraints required by \cref{lst:explore_path} are covered by each approach.}\label{tab:existing_constraints}
\renewcommand{\arraystretch}{1.0}  
\resizebox{1.0\linewidth}{!}{
\begin{threeparttable}
\begin{tabular}{l|l|l|l|l|l|l|l}  
\toprule
\multicolumn{1}{c|}{\multirow{2}{*}{Tool}} & \multicolumn{1}{c|}{\multirow{2}{*}{Extracted Constraints}}  & \multicolumn{5}{l|}{Constraints Required By~\cref{lst:explore_path}} \\ \cmidrule{3-7}
                      &                                         &   SC1 & TB1 & FB1 & SC2 & SC3 \\ \midrule
TensorScope & 1: {\mycode{}diag\_index.num\_element > 0}              & \ding{52}      & \ding{55}      & \ding{55}      & \ding{55}     & \ding{55}     \\ \midrule
DocTer*     & 
\begin{tabular}[c]{@{}l@{}}
1: {\mycode{}num\_cols\_tensor.default\_value==-1 \&\& num\_rows\_tensor.default\_value==-1}
\\ \quad{\mycode{}\&\& num\_cols\_tensor.range==[0, inf] \&\& num\_rows\_tensor.range==[0, inf]}
\end{tabular}
& \ding{55}              & \ding{55}              & \ding{55}              & \ding{52}             & \ding{52}             \\ \midrule
ACETest     &
\begin{tabular}[c]{@{}l@{}}
1: {\mycode{}diag\_index.shape[0] > 1 \&\& diag\_index.num\_element > 0 \&\& \dots}\\
2: {\mycode{}diag\_index.shape[0] <= 1 \&\& diag\_index.num\_element > 0 \&\& \dots}\\
\dots
\end{tabular} 
& \ding{52}              & \ding{52}              & \ding{52}              & \ding{55}             & \ding{55}             \\ \midrule
\toolname{} &
\begin{tabular}[c]{@{}l@{}}
1: {\mycode{}diag\_index.shape[0] == 1 \&\& diag\_index.num\_element > 0 \&\& \dots}\\
2: {\mycode{}diagonal.shape[-1] >= 0  \&\& diagonal.ndims >= 1}\\
3: {\mycode{}diagonal.shape==[3] \&\& diag\_index==1 \&\& num\_rows\_tensor==num\_cols\_tensor==4}\\
\dots
\end{tabular}
& \ding{52}              & \ding{52}              & \ding{55}              & \ding{52}             & \ding{52}             \\ \midrule
\end{tabular}
\begin{tablenotes}
\item [1] Since DocTer does not cover the API {\mycode{}tf.raw\_ops.MatrixDiagV2}, we use its constraint on a similar API {\mycode{}tf.compat.v1.matrix\_diag} as a reference. These two APIs share similar implementations and functionalities.\\
\item [2] `SC' is short for `Sanity Check'; `TB' is short for `True Branch'; `FB' is short for `False Branch'.
\end{tablenotes}
\end{threeparttable}
}
\end{table}

Table~\ref{tab:existing_constraints} demonstrates the limitation of existing approaches~\cite{docter,tensorscope,acetest} in extracting input constraints.
For the code snippet in Listing~\ref{lst:explore_path}, TensorScope can extract the constraint for sanity check 1 and DocTer can only cover part of the constraints (\ie, {\mycode{}num\_rows\_tensor==-1} and {\mycode{}num\_rols\_tensor==-1}) required by sanity check 2 and 3.
Both tools extract no constraints for the {\mycode{}True/False} branches of the {\mycode{}If} statement at line 4.
As for ACETest, although it can extract constraints for sanity check 1 at line 3 and {\mycode{}True/False} branches of the {\mycode{}If} statement at line 4, it still misses constraints required by sanity checks 2 and 3 (lines 13--14).
This is because these sanity checks involve local variables {\mycode{}min\_num\_rows} and {\mycode{}min\_num\_cols}, which are defined using external functions {\mycode{}std::min} and {\mycode{}std::max}, and these external functions are neither comprehended by existing solvers nor analyzed by ACETest.
Since DL library APIs often rely on external libraries or modules, external functions are frequently used in the source code of these APIs.
Simply skipping the analysis of these external functions can cause some input constraints required by execution paths not extracted by static analysis, resulting in incomplete path constraints.
However, due to the large codebase of DL libraries and the inherent scalability challenge of static analysis~\cite{scale_symbolic_execution, scalable_symbolic_execution_2011}, it is prohibitively expensive to perform static analysis on all external functions.
Efficiently analyzing these external functions for path constraint extraction in DL library APIs remains challenging.

\subsection{Opportunities and Challenges}\label{subsec:opportunities_and_challenges}
With the emergence of LLMs and their demonstrated domain knowledge on the usage and implementation of DL library APIs~\cite{titanfuzz,fuzzgpt}, one may wonder if we can apply LLMs for DL library counterpart identification and path constraint extraction
However, our initial attempts to use LLMs for these tasks have shown limited success. 
Merely prompting LLMs to synthesize counterparts and extract path constraints is ineffective, as demonstrated in the following discussion.

\begin{listing}
\begin{minted}[
    baselinestretch=1.0,
    fontsize=\scriptsize,
    xleftmargin=0.5ex,
    bgcolor=bg,
    breaklines=true,
    escapeinside=||,
]{text}
|\underline{\textbf{Basic Prompt}}|:
Find an counterpart in PyTorch to implement the same functionality of this 
TensorFlow API: 
tf.raw_ops.SpaceToBatchND(input=input,block_shape=block_shape, paddings=paddings)
|\underline{\textbf{Counterpart Generated By This Basic Prompt}}|
def space_to_batch_nd(input, block_shape, paddings):
  if len(block_shape) != input.dim() - 2:
      raise ValueError("Block shape must match the number of spatial dimensions.")
  ... // omitted 
  reshaped = padded_input.view(batch_size, channels, *padded_shape)
  # Permute to get the desired order
  permuted = reshaped.permute(0, 2, 3, *range(4, reshaped.dim()))
  return permuted
|\underline{\textbf{Example Input That Exposes Functionality Discrepancies Between The Target API and Its Counterpart}}|
block_shape = tf.constant([2, 1], dtype=tf.int32)
input = tf.constant([[[1],[2]],[[3],[4]]], dtype=tf.float32)
|\underline{\textbf{Target API's Result}}|
[[[1.]],[[3.]],[[2.]],[[4.]]]
|\underline{\textbf{Counterpart's Result}}|
|\textcolor{bgRed}{\textbf{Exception with error message: Block shape must match the number of spatial dimensions.}}|
\end{minted}
\caption{Counterpart synthesized via a basic prompt}
\label{lst:simple_counterpart}
\end{listing}

\textit{Merely prompting LLMs to synthesize counterparts is ineffective.} Listing~\ref{lst:simple_counterpart} illustrates a basic prompt for synthesizing a counterpart with the \llm{} model.
While the LLM can return a counterpart for API {\mycode{}tf.raw\_ops.SpaceToBatchND}, it shows two problems. 
First, since LLM's generation result could be wrong, one problem is: \textit{how to verify the validity of synthesized counterparts?}
Indeed, our analysis reveals that this synthesized counterpart can be invalid under certain input conditions.
For instance, when the input argument {\mycode{}input} is a 3D tensor and {\mycode{}block\_shape} is a 1D tensor containing two elements (see the example input in Listing~\ref{lst:simple_counterpart}), the target API returns a meaningful result while the counterpart directly raises an exception.
When using this counterpart for differential testing, false positives could be introduced if test inputs fall into this input condition.
Therefore, to reduce false positives, synthesized counterparts are expected to be valid for various test inputs, thus introducing the second problem: \textit{how to improve the quality of synthesized counterparts so they can be valid for various test inputs?}

\begin{listing}
\begin{minted}[
    baselinestretch=1.0,
    fontsize=\scriptsize,
    xleftmargin=0.5ex,
    bgcolor=bg,
    breaklines=true,
    escapeinside=??,
]{text}
?\underline{\textbf{Basic Prompt}}?:
Given the following implementation of API: tf.raw_ops.MatrixDiagV2(diagonal, k, num_rows, num_cols, ...
For each execution path in the below code snippet, generate path constraint on tf.raw_ops.MatrixDiagV2's input arguments to reach that execution path.
...

?\underline{\textbf{LLM's Output For This Basic Prompt}}?
Path 1: Single Element in diag_index
- diag_index.shape() must be a scalar or vector
- diag_index.NumElements() == 1   ---->  ?\textbf{This constraint can guide test input to pass Sanity Check 1}?
...
Path 2: Two Elements in diag_index
- diag_index.shape() must be a scalar or vector
...
Path 3: num_rows is a valid scalar
- num_rows_tensor.shape() must be a scalar
...
Path 4: num_cols is a valid scalar
- num_cols_tensor.shape() must be a scalar
...
Path 5: Valid diagonal shape
- diagonal.shape() must be at least 1-dim
...
Path 6: Validate minimum rows and columns
- num_rows == -1 || num_rows >= min_num_rows
- num_cols == -1 || num_cols >= min_num_cols
Path 7: Square output assumption
- num_rows == -1 && num_cols == -1   ---->  ?\textbf{This constraint can guide test input to pass the Sanity Check 2,3, but \textcolor{bgRed}{test inputs' diversity is limited since num\_rows and num\_cols can only be -1}}?
\end{minted}
\caption{Path constraints synthesized via a basic prompt. We highlight useful constraints that can guide test input generation to pass related sanity checks or branches. The LLM's output is simplified.}
\label{lst:simple_constraints}
\end{listing}
\textit{Merely prompting LLMs to extract path constraints is ineffective.}
When directly prompting an LLM to analyze DL library APIs' implementation for path constraint extraction, we observe that the LLM's effectiveness is also limited.
For instance, we directly send the implementation of {\mycode{}tf.raw\_ops.MatrixDiagV2} to the \llm{} with a basic prompt, requesting it to extract path constraints for the code in Listing~\ref{lst:explore_path}. 
When analyzing the LLM's response, we observe that although seven path constraints can be generated by \llm{}, most of them are incomplete (see Listing~\ref{lst:simple_constraints}).
In particular, only two path constraints (\ie, Path 1 and Path 7) can guide test input to pass one or two sanity checks.
Moreover, some path constraints may harm the test diversity by restricting input arguments to specific values (\eg, Path 7 restricts the {\mycode{}num\_rows} and {\mycode{}num\_cols} to -1, prohibiting the generation of other valid values for these input arguments).
\textit{One possible reason is the large codebase of DL library API's implementation hinders the LLM's performance in extracting path constraints.}
For example, the implementation of {\mycode{}tf.raw\_ops.MatrixDiagV2} has 120 lines, including Python and C++ code.
Analyzing the entire implementation directly to extract path constraints for specific execution paths may be challenging for the LLM\@.

\section{Methodology}

\begin{figure*}[t]
    \centering
    \includegraphics[width=\textwidth]{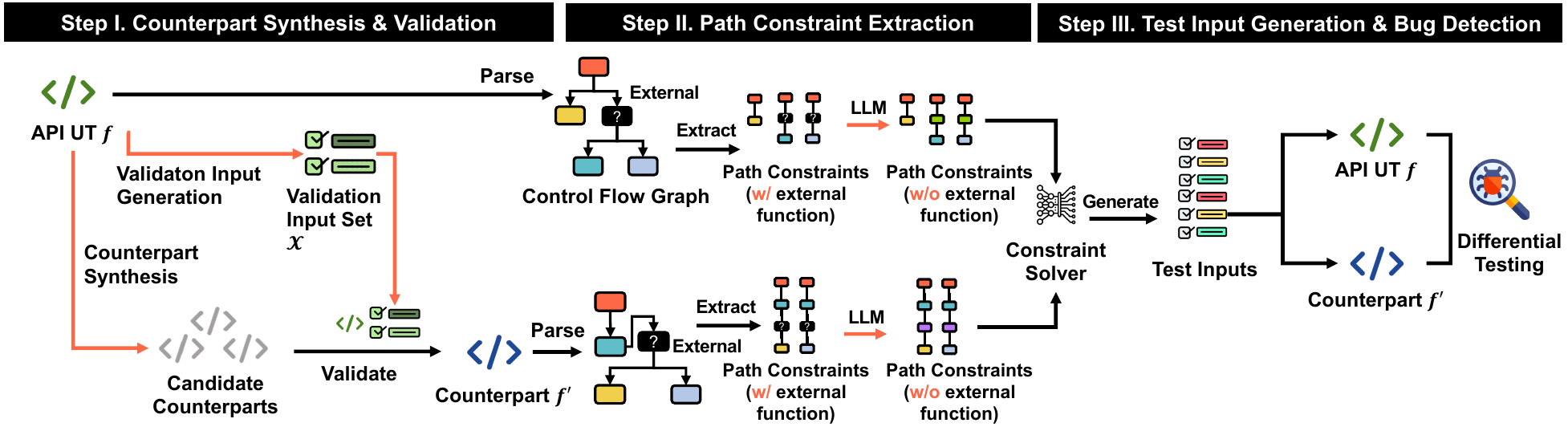}
    \caption{Workflow of \toolname{} with three steps: counterpart synthesis and validation (\cref{subsec:counterpart_synthesis}), path constraint extraction (\cref{subsec:method_path_constraint}), and test input generation \& bug detection (\cref{subsec:method_input_generation})}.
    \Description{Workflow of \toolname}\label{fig:methodology_overview}
\end{figure*}

Figure~\ref{fig:methodology_overview} shows the workflow of \toolname.
For each DL library API under test \undertest, \toolname{} will apply the following three steps:

\textit{Step I. Counterpart Synthesis and Validation (\cref{subsec:counterpart_synthesis})}. 
\toolname{} first leverages the LLM to synthesize a counterpart \counterpart{} that simulates the intended functionality of the API \undertest{} by making use of APIs from a different DL library.
Since LLM may make mistakes, \toolname{} validates each synthesized counterpart on a diverse set of validation inputs \validinputs{} generated by \cref{subsubsec:validation_input_generation}.
The synthesized counterpart is considered to be valid over \validinputs{} if it does not crash and produces consistent output with \undertest{} on every input in \validinputs{}.

\textit{Step II\@. Path Constraint Extraction (\cref{subsec:method_path_constraint})}. 
After synthesizing the counterpart \counterpart, we further extract the path constraints in both \undertest{} (the API under test) and \counterpart{} (the corresponding counterpart). 
Note that we only focus on the constraints that confine the input arguments because these extracted constraints will finally serve as guidance to generate test inputs (Step III). 
Additionally, \undertest{} and \counterpart{} often invoke external functions (as motivated in \cref{subsec:limitation_constraint}), so we also turn to the LLM to infer constraints from paths involving external functions. 
By doing so, we can eliminate the extensive static analysis on external functions and their dependencies while extracting as many path constraints from \undertest{} and \counterpart{} as possible.

\textit{Step III\@. Test Input Generation and Bug Detection (\cref{subsec:method_input_generation})}.
In this step, \toolname{} iteratively generates test inputs based on the extracted path constraints.
In each iteration, one path constraint is selected to generate a test input, \toolname{} further tests \undertest{} and \counterpart{} with this input, and employs differential testing to detect output inconsistencies.

\subsection{Step I. Counterpart Synthesis and Validation}\label{subsec:counterpart_synthesis}
\textbf{Criterion of Valid Counterparts}.
Following existing approaches~\cite{deeprel,numerical_discrepancies_issta}, we consider a counterpart \counterpart{} of the target API \undertest{} to be \textit{valid} over \textit{validation inputs} \validinputs{} if they satisfy Equation~\ref{eq:counterpart_definition}, where $\epsilon$ is a small positive value (\eg, 0.1).

\begin{equation}
\forall x \in \mathcal{X}, |f(x)-f'(x)|\leq\epsilon
\label{eq:counterpart_definition}
\end{equation}

Although it is straight-forward to ask the LLM to synthesize a counterpart simulating \undertest{}, there are two problems remaining addressed. 
First, \textit{how to validate the synthesized counterparts?} LLMs can make mistakes, while it is non-trivial to distinguish valid ones from invalid ones due to diverse implementations.
To address this problem, we generate a diverse set of validation inputs (\ie, \validinputs{}) to verify the validity of synthesized counterparts through Equation~\ref{eq:counterpart_definition}.
This set of validation inputs is produced by initially gathering inputs from an LLM-based generation strategy as proposed in a latest work~\cite{titanfuzz}, and then diversifying them using a property mutation strategy (\cref{alg:valid_input_generation}).
Yet, it is not easy for LLMs to synthesize valid counterparts for some APIs (as demonstrated in \cref{subsec:opportunities_and_challenges}).
Thus, the second problem is: \textit{how to guide LLMs to synthesize valid counterparts effectively?}
To address this problem, \toolname{} proposes a counterpart synthesis approach (\cref{subsubsec:counterpart_synthesis}) that guides LLM to iteratively synthesize valid counterparts over \validinputs{}.

\subsubsection{Validation Input Generation}\label{subsubsec:validation_input_generation}
Due to the large input space of API under test \undertest, it is hard to enumerate all input values to validate a synthesized counterpart.
Therefore, a finite set of validation inputs is commonly constructed to validate synthesized counterparts~\cite{deeprel, numerical_discrepancies_issta}. 
\toolname{} constructs a finite set of validation inputs in two stages: \textbf{{LLM Prompting}} (leveraging an existing approach to prompt an LLM for initial validation input set) and \textbf{{Property Mutation}} (diversifying the validation input set via property mutation). We explain the details as follows.

\textbf{Stage 1. LLM Prompting}. 
In this stage, \toolname{} adopts a typical validation input generation strategy proposed by TitanFuzz~\cite{titanfuzz}, which prompts an LLM using zero-shot.
More specifically, the prompt instructs LLM step-by-step by specifying the target library (\eg, TensorFlow) and the API signature, with concrete task instruction (see \cref{lst:prompt_direct_prompting}).
However, inputs generated by the same prompt are likely to have similar characteristics, such as identical tensor shapes or values~\cite{fuzzgpt}. 
In addition, directly prompting the LLM to generate diverse inputs may not be sufficient due to the lack of clear criteria defining diversity.
Consequently, inputs generated via LLM prompting may lack the necessary diversity to validate synthesized counterparts.
To address this limitation, \toolname{} further diversifies generated validation inputs in stage 2 via property mutation.

\begin{listing}[htbp]
\begin{minted}[
    baselinestretch=1.0,
    fontsize=\scriptsize,
    xleftmargin=0.5ex,
    bgcolor=bg,
    breaklines=true,
    escapeinside=||,
]{text}
|\underline{\textbf{Prompt}}|:
Task 1: Import TensorFlow 2.10.0
Task 2: Generate valid parameter... The name of parameter variables should be: `image', ...
Task 3: Call the function: tf.image.pad_to_bounding_box(image, ...)
\end{minted}
\caption{An example prompt used by LLM prompting. The prompt is simplified.}
\label{lst:prompt_direct_prompting}
\end{listing}

\textbf{Stage 2. Property Mutation.}
Stage 2 targets diversifying validation inputs for counterpart \counterpart{}.
A validation input typically consists of multiple arguments.
For each argument, \toolname{} follows COMET~\cite{comet} to diversify its properties, such as {\mycode{}shape} for tensor arguments and {\mycode{}value} for other primitive arguments.
Specifically, given an argument, \toolname{} produces a set of mutants by enumerating all possible property values within the mutation space of each property.
Mutation space for each argument property is manually defined in \cref{tab:api_properties}.
Since the mutation space for non-continuous arguments (\eg, {\mycode{}float}) could be large, it is hard to enumerate all possible values in this space.
Consequently, \toolname{} follows COMET~\cite{comet} by generating five mutants for each non-continuous argument (\eg, five distinct float values will be generated for a float argument).
Additionally, for \texttt{string} arguments or arguments in other types, \toolname{} does not apply property mutation on these arguments since their mutation spaces are often varied across different APIs.
After generating mutants by property mutation, \toolname{} further examines the validity of these mutants via executing \undertest{} on these mutants.
Invalid mutants will be filtered out since they cannot validate the output correctness of synthesized counterparts via Equation~\ref{eq:counterpart_definition}.

\begin{table}[t!]
\caption{DL library API argument properties and their mutation spaces.}\label{tab:api_properties}
\renewcommand{\arraystretch}{0.9}  
\setlength{\tabcolsep}{2.2pt}
\resizebox{1\linewidth}{!}{
\begin{tabular}{l|l|l|l}  
\toprule
Argument Type   &   Property & Description &  Mutation Space \\ \midrule
\multirow{4}{*}{Tensor} &   ndims    & integer, number of dimensions & [0, 5] \\ \cmidrule{2-4}
                        &   shape[i] & integer, number of elements on dimension i & [1, 5] \\ \cmidrule{2-4}
                        &   dtype    & string, data type &
                                    \begin{tabular}[c]{@{}l@{}}
                                    integer tensor: `int32', `int64'\\
                                    float tensor: `float16', `float32', `float64'\\
                                    complex tensor: `complex64', `complex128'
                                    \end{tabular} \\ \cmidrule{2-4}
                        &   num\_element & integer, number of elements & [1, $\infty$) \\ \midrule
Bool                    &   value    & boolean & \{True, False\} \\ \midrule
Integer                 &   value    & integer & [-5, 5]     \\ \midrule
Float                   &   value    & integer & [-100, 100]     \\ \midrule
String                  &   value    & string & N/A     \\ \midrule
Others                  &   value    & other types & N/A     \\
\bottomrule
\end{tabular}
}
\end{table}

\begin{algorithm}[htbp]
\caption{Validation Input Generation}\label{alg:valid_input_generation}
\LinesNumbered{}
\SetKwProg{Def}{def}{:}{}
\SetKwData{llminputs}{$\mathcal{X}_{llm}$\xspace}
\SetKwData{mutant}{$\mathcal{X}_{mutant}$\xspace}
\SetKwData{numllminputs}{\ensuremath{N}\xspace}
\SetKwData{singleinput}{$i$\xspace}
\SetKwData{repairedinput}{$i'$\xspace}
\SetKwData{seedinput}{$i_{seed}$\xspace}
\SetKwData{argument}{$\arg$\xspace}
\SetKwData{valuespace}{$Space_{value}$\xspace}
\SetKwData{value}{$value$\xspace}
\SetKwData{property}{$property$\xspace}
\KwIn{
    \undertest, the target API whose validation inputs need to be generated;
}
\KwIn{
    \numllminputs, the required number of inputs generated by the LLM\@;
}
\KwOut{\validinputs, the validation input set}
\Def{ValidationInputGeneration(\undertest, \numllminputs)}{   
    \validinputs $\leftarrow \varnothing$ \\   
    \llminputs $\leftarrow$ sampling \numllminputs inputs via \underline{\textit{Stage 1.LLM Prompting}} \\   
    \ForEach{\singleinput in \llminputs}{   
        \If{IsValidInput(\undertest, \singleinput)}{   
            \validinputs.add(\singleinput)   
        }
    }
    // \underline{\textit{Stage 2.Property Mutation}}  \\
    \seedinput $\leftarrow$ sample \validinputs \\  
    \ForEach{\argument in \seedinput}{  
        \validmutants $\leftarrow$ PropertyMutation(\seedinput, \argument)\\   
        \validinputs $\leftarrow$ \validinputs $\cup$ \validmutants  
    }
    \textbf{return \validinputs}
}

\Def{PropertyMutation(\seedinput, \argument, \undertest)}{  
    \mutants $\leftarrow \varnothing$  \\  
    \ForEach{\property in \argument}{  
        \valuespace $\leftarrow$ extract \property's mutation space from Table~\ref{tab:api_properties}\\   
        \ForEach{\value in \valuespace}{  
            $\mutant \leftarrow$ change the \seedinput's \underline{\property} to \underline{\value} \\
            \If{IsValidInput(\undertest, \mutant)}{   
                \mutants.add(\mutant)  
            }
        }
    }
    \textbf{return \mutants}
}
\end{algorithm}

Algorithm~\ref{alg:valid_input_generation} illustrates the detailed process of validation input generation.
For each API \undertest, \toolname{} first uses LLM prompting (line 3) to sample \textit{N} (\textit{N}=3) inputs, adding those valid to the validation input set \validinputs{} (lines 4--6).
Based on the validation input set generated from the LLM prompting, \toolname{} applies property mutation to enumerate different property values for each argument of \undertest{} (lines 13--21).
Those valid mutants will be added to \validinputs{} (line 11).
The final validation inputs \validinputs{} contain valid inputs and mutants from both LLM prompting and property mutation.
If no validation input is generated (\ie, \validinputs{}$=\varnothing$), \toolname{} skips the testing of this API because no counterpart can be reliably validated through Equation~\ref{eq:counterpart_definition}.

\subsubsection{Counterpart Synthesis}\label{subsubsec:counterpart_synthesis}
To synthesize a counterpart \counterpart{} for an API \undertest{}, \toolname{} instructs the LLM with a prompt including this API's signature and example inputs sampled from previously generated validation input set \validinputs{}.
The role of example inputs and API's signature is guiding the LLM to synthesize a function with the same signature and also supporting the same test inputs as the given API\@.
\cref{lst:prompt_counterpart_collection} exemplifies the prompt as well as the LLM's response when \toolname{} queries this LLM to synthesize a counterpart for API {\mycode{}tf.scatter\_nd}.
In this prompt, \toolname{} includes three example inputs sampled from \validinputs{} and associates these inputs with the given API's signature, instructing the LLM to search for an equivalent function.
In particular, \toolname{} restricts the LLM to synthesize a function using APIs from a different DL library (\eg, PyTorch), to avoid searching for the given API's caller, callee, or alias within the same DL library.
Each synthesized counterpart will be further validated on \validinputs{} by comparing its outputs with its target \undertest{} via Equation~\ref{eq:counterpart_definition}.
Once a synthesized counterpart is identified invalid, \toolname{} uses its comparison result on \validinputs{} to construct a feedback prompt, instructing the LLM to refine the original counterpart.
We next discuss the details of our feedback prompt.

\begin{listing}[htbp]
\begin{minted}[
    baselinestretch=1,
    fontsize=\scriptsize,
    xleftmargin=0.5ex,
    bgcolor=bg,
    breaklines=true,
    label={lst:prompt_counterpart_collection},
    escapeinside=||
]{text}
|\underline{\textbf{Prompt}}|:
Given the following sample inputs, your task is to generate a function named `pytorch_call` that uses pytorch's API.
The generated function SHOULD have the same functionality as the `tensorflow_call` ...

|\textit{[Sample Inputs]}|
input 1: indices = tf.constant([[0], [2]]);updates=...;shape=...
input 2: ...
input 3: ...
|\textit{[Function using TensorFlow API]}|
def tensorflow_call(indices,updates,shape):
    return tf.scatter_nd(indices,updates,shape)
|\textit{[Function using PyTorch APIs]}|
def pytorch_call(indices,updates,shape):
|\underline{\textbf{LLM}}:|
def pytorch_call(indices,updates,shape):
    ...
    tensor_shape=torch.Size(shape)
    output_tensor=torch.zeros(tensor_shape)
    |\underline{\textbf{\textcolor{bgRed}{output\_tensor.scatter\_(0, tensor\_indices,tensor\_updates)}}}| |\textbf{\textcolor{bgRed}{// this line is incorrect!}}|
    return output_tensor
\end{minted}
\caption{An example prompt for counterpart synthesis. The prompt is simplified. This synthesized counterpart is invalid because the \underline{underlined line} is incorrect.}
\label{lst:prompt_counterpart_collection}
\end{listing}

Given the comparison result of an invalid \counterpart{}, \toolname{} constructs its feedback prompt by including the details of crash and inconsistency cases in this comparison result.
As is exemplified in Listing~\ref{lst:prompt_counterpart_feedback}, each crash case in the prompt contains the input (denoted as ``[\underline{Crash on Input}]'') and related crash error message (denoted as ``[Error Message]'').
As for the inconsistency case, the prompt includes the input (denoted as ``[\underline{Inconsistent on Input}]''), output from API under test \undertest{} (denoted as ``[Expected Output]''), and output from the \counterpart{} (denoted as ``[Actual Output]'').
To prevent exceeding the LLM's token limit, \toolname{} selectively includes crash cases with unique error messages and one randomly selected inconsistency case in the prompt.
Leveraging this feedback prompt, the LLM successfully refines the originally invalid counterpart to a valid one (see the LLM's output in Listing~\ref{lst:prompt_counterpart_feedback}).

\begin{listing}[htbp]
\begin{minted}[
    baselinestretch=1,
    fontsize=\scriptsize,
    xleftmargin=0.5ex,
    bgcolor=bg,
    breaklines=true,
    label={lst:prompt_counterpart_collection},
    escapeinside=||
]{text}
|\underline{\textbf{Follow-up Prompt of Listing \ref{lst:prompt_counterpart_collection}}}|:
The generated function crashes on the following input:
[|\underline{Crash on Input}|]:
indices = tf.constant([[0], [2]])...
[Error Message]:
"Index" tensor must have the same number of dimentions as "self" tensor...

The generated function produces incorrect output on the following input:
[|\underline{Inconsistent on Input}|]:
indices = tf.constant([[0], [2]])...
[Expected Output]:
...
[Actual Output]:
...

|\underline{\textbf{LLM}}|:
def pytorch_call(...):
    ...
    |\textbf{\textcolor{bgGreen}{output\_tensor.scatter\_(dim=0, index=tensor\_indices.squeeze(), src=tensor\_updates)}}| |\textbf{\textcolor{bgGreen}{// the refined line is correct}}|
    return output_tensor
\end{minted}
\caption{An example of a feedback prompt that guides the LLM to refine an invalid counterpart based on its comparison result on a validation input set.}
\label{lst:prompt_counterpart_feedback}
\end{listing}

\begin{algorithm}[htbp]
    \caption{Iterative Counterpart Synthesis With Execution Feedback}\label{alg:counterpart_synthesis}
    \LinesNumbered{}
    \SetKwProg{Def}{def}{:}{}
    \KwIn{
        \undertest, a target API whose counterpart needs to be synthesized;
    }
    \KwIn{
        \validinputs, a set of validation inputs
    }
    \KwOut{\counterpart, the counterpart satisfying Equation~\ref{eq:counterpart_definition}}
    \SetKwData{execution}{$property$\xspace}
    \SetKwData{allcase}{$S_{X}$\xspace}
    \SetKwData{crashcase}{$C_{crash}$\xspace}
    \SetKwData{unicracase}{$C_{unique\_crash}$\xspace}
    \SetKwData{inconcase}{$C_{incon}$\xspace}
    \SetKwData{consistentcase}{$C_{consnt}$\xspace}
    \SetKwData{case}{$c_{incon}$\xspace}
    \Def{CounterpartSynthesis(\undertest, \validinputs)}{   
        \counterpart $\leftarrow$ synthesize a counterpart via LLM{}   // Listing~\ref{lst:prompt_counterpart_collection}  \\   
        \crashcase, \inconcase $\leftarrow$ validate \counterpart on \validinputs  \\  
        \If{\crashcase $== \varnothing$ $\land$ \inconcase $== \varnothing$}{  
            \textbf{return \counterpart}
        }
        iteration $\leftarrow 1$ \\
        \While(){$\neg$ (\crashcase $== \varnothing$ $\land$ \inconcase $== \varnothing$) $\land$ iteration <= 5}{   
            \case $\leftarrow$ sample one inconsistency case from \inconcase  \\  
            \unicracase $\leftarrow$ get unique crash cases from \crashcase  \\  
            \counterpart $\leftarrow$ refine \counterpart{} via LLM with \unicracase and \case \space\space\space// Listing~\ref{lst:prompt_counterpart_feedback}  \\   
            \crashcase, \inconcase $\leftarrow$ validate \counterpart on \validinputs  \\  
            iteration += 1\\
        }
        \If{\crashcase $== \varnothing$ $\land$ \inconcase $== \varnothing$}{  
            \textbf{return \counterpart}
        }
    }
    \end{algorithm}

Algorithm~\ref{alg:counterpart_synthesis} illustrates the counterpart synthesis process in detail.
Given an API under test \undertest{} and its validation inputs \validinputs{}, \toolname{} first leverages an LLM to synthesize a counterpart \counterpart{} using a counterpart synthesis prompt exemplified in Listing~\ref{lst:prompt_counterpart_collection} (line 3).
A counterpart \counterpart{} is deemed valid if no \crash{} and \inconsistent{} case is found on any input in \validinputs{} (lines 3--5).
Otherwise, \toolname{} queries the LLM with the feedback prompt (exemplified in Listing~\ref{lst:prompt_counterpart_feedback}), instructing it to refine the invalid counterpart (lines 8--10).
\toolname{} iteratively refines this invalid counterpart until either a correct counterpart is produced or the iteration limit (five iterations) is reached.
To increase the chance of finding a valid counterpart, the Algorithm~\ref{alg:counterpart_synthesis} is repeated three times for each API under test.

\subsection{Step II. Path Constraint Extraction}\label{subsec:method_path_constraint}
For a DL library API \undertest{} and its valid counterpart \counterpart{} synthesized in \cref{subsec:counterpart_synthesis}, \toolname{} next extracts path constraints for each of them individually.
Path constraints are extracted via analyzing each execution path from the control flow graph of the source code.
Since these path constraints are further served as the guidance for test input generation in \cref{subsec:method_input_generation}, 
\toolname{} only extracts constraints for API inputs (referred to as input constraints) required by execution paths during the analysis.
Each path constraint is formatted as a conjunction of input constraints required by an execution path.
To construct a path constraint, \toolname{} traverses concerned execution path, tracking API inputs along it, and finally constructs input constraints from certain statements (\cref{subsubsec:constraint_construction}) in this path.
However, if an constructed input constraint involves external functions, whose behaviors are unknown to SMT solvers, the constructed constraint is unsolvable.
Therefore, \toolname{} further proposes an input constraint inference technique (\cref{subsubsec:input_constraint_inference}), leveraging an LLM to infer solvable input constraints from those unsolvable ones.
We next present the detailed constraint extraction process for each path.

\subsubsection{Constraint Construction}\label{subsubsec:constraint_construction}
When constructing input constraints from execution paths, we focus on two primary types of statements: conditional statements (\eg, \texttt{if}) and sanity check statements, which refer to common function calls used for validation (\eg, the \texttt{OP\_REQUIRES} function call as seen in line 3 of Listing~\ref{lst:explore_path}).
As is demonstrated by existing works~\cite{tensorscope, acetest}, input constraints are commonly encoded within these statements, and test inputs failing to comply with these constraints are impeded from reaching further within the path.
After identifying these statements, the next step is constructing input constraints from them.
To do so, \toolname{} first extracts conditions from these statements, including branch conditions from conditional statements such as \texttt{if} statements, and conditions extracted from sanity check statements via specific rules.
Inspired by existing works~\cite{tensorscope, acetest}, we have manually implemented rules for each type of sanity check statement to extract conditions, which are summarized in Table~\ref{tab:sanity_check_constraints}.
For instance, when handling a sanity check statement calling the function \texttt{OP\_REQUIRES}, we extract its second argument as the condition. 

\begin{figure}[htbp]
\centering
\includegraphics[width=\linewidth]{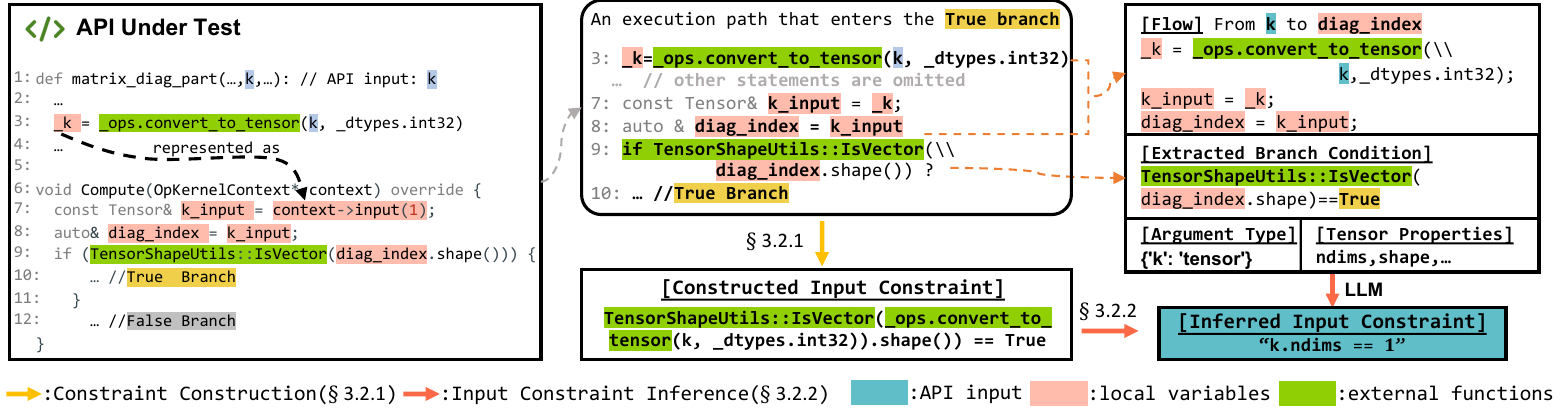}
\caption{A running example of \toolname{} that extracts an input constraint from the {\mycode{}if} statement (line 9) of an execution path. \toolname{} first constructs input constraint via \S~\ref{subsubsec:constraint_construction}. 
Since the constructed input constraint involves external functions, which are unsolvable to SMT solvers, \toolname{} continues \S~\ref{subsubsec:input_constraint_inference} to infer input constraint via an LLM.
The prompt used in \S~\ref{subsubsec:input_constraint_inference} includes (a) the flow from the API input (\texttt{k}) to the variable (\texttt{diag\_index}) involved in the \texttt{if} statement, (b) branch condition extracted from the \texttt{if} statement, (c) argument type of API input, and (d) candidate tensor properties.}\label{fig:path_constraint_demonstration}
\Description{A running example of \toolname{} that uses the LLM to extract valid input constraints from a conditional statement with external functions.}
\end{figure}

Once conditions are extracted, \toolname{} further constructs input constraints from them.
This step is crucial as conditions extracted from concerned statements may not directly relate to API inputs.
Taking Figure~\ref{fig:path_constraint_demonstration} as an example, the extracted branch condition of \texttt{if} statement at line 9 does not involve any API input.
Instead, this condition involves a local variable, \texttt{diag\_index}.
Directly using this condition for test input generation is not helpful since solving it cannot provide guidance for generating API inputs.
To construct input constraints from conditions involving local variables, \toolname{} follows ACETest~\cite{acetest} to traverse the execution path.
This traversal tracks the flow of API inputs by analyzing all declaration and assignment statements along this path.
The rationale is to identify which variables are related to API inputs.
A variable is considered related if declared or assigned based on either API inputs or other related variables in the execution path.
Once all related variables are identified, \toolname{} checks whether the variables involved in the extracted conditions are among them. 
If no related variables are found, the extracted condition is deemed irrelevant to API inputs, and no input constraint will be derived from it.
Conversely, if related variables are found, \toolname{} converts the extracted condition into an input constraint based on the flow from API inputs to the related variables (see [\underline{Constructed Input Constraint}] in Figure~\ref{fig:path_constraint_demonstration}).

\subsubsection{Input Constraint Inference}\label{subsubsec:input_constraint_inference}
However, these constructed input constraints may involve functions written in external libraries (\eg, Eigen, a popular C++ library for linear algebra) or external modules (\eg, TensorUtils).
Taking the \texttt{if} statement at line 9 of Figure~\ref{fig:path_constraint_demonstration} as an example, the constructed constraint for input \texttt{k} (see [\underline{Constructed Input Constraint}] in Figure~\ref{fig:path_constraint_demonstration}) involves two external function calls, \texttt{convert\_to\_float} and \texttt{IsVector}.
The former is introduced during the flow from \texttt{k} to \texttt{diag\_index}, while the latter is from the branch condition of the \texttt{if} statement.
Due to the large codebase of these external functions and their dependencies, it is hard to analyze all external function calls when extracting input constraints.
Consequently, external functions are inevitable in some extracted input constraints, making these constraints unsolvable to SMT solvers.
To handle constraints with external functions, existing works~\cite{tensorscope,acetest} either skip these constraints or manually model the behavior only on a limited set of external functions.
Both ways may lead to incomplete path constraints.

We propose an LLM-based input constraint inference strategy to handle constraints with external functions.
The insight behind using LLM is that \textit{LLMs have learned domain-specific knowledge about DL libraries and their upstream libraries, enabling the analysis of external functions even without their source code}.
However, as demonstrated in \cref{subsec:opportunities_and_challenges}, the effectiveness of LLM in extracting constraints may be limited when the source code is large.
To reduce the size of the source code provided to LLM, \toolname{} handles each input constraint with external functions individually by providing the following four elements to guide the LLM in inferring a valid input constraint:
\begin{enumerate}[topsep=0pt, leftmargin=*]
\item Extracted condition. This part includes the condition extracted from a sanity check or conditional statement.
\item Flow from API inputs to related local variables. We provide the LLM with the flow from API inputs to related local variables involved in the extracted condition.
\item Argument types. To facilitate LLM in structuring the input constraint appropriately, we provide the argument types of API inputs.
\item Tensor properties. This part is particularly designed for tensor inputs since tensor constraints primarily target specific tensor properties~\cite{acetest}. 
By providing candidate tensor properties, we follow ACETest to restrict the tensor constraint, focusing on these specific properties.
This part is omitted for non-tensor inputs.
\end{enumerate}

As is exemplified in Figure~\ref{fig:path_constraint_demonstration}, since the input constraint constructed from the \texttt{if} statement at line 9 involves external functions, \toolname{} utilizes the above four elements to construct a prompt (see Listing~\ref{lst:methodology_input_condition_prompt}), instructing the LLM to infer a valid input constraint.
This prompt includes the extracted condition (\underline{[Condition]}), the flow from API inputs to related local variables (\underline{[Execution Trace]}), the argument type of API input {\mycode{}k} and the tensor properties since {\mycode{}k} is a tensor.
Based on this constructed prompt, despite the original constraint involves external functions like {\mycode{}convert\_to\_float} and {\mycode{}IsVector}, the LLM effectively derives the correct input constraint, which is {\mycode{}k.ndims == 1}.  

\begin{listing}[htbp]
\begin{minted}[
    baselinestretch=1.0,
    fontsize=\scriptsize,
    xleftmargin=0.5ex,
    bgcolor=bg,
    breaklines=true,
    escapeinside=||
]{text}
|\underline{\textbf{Prompt}}|:
Here is one execution path
|\textbf{[Execution Trace]}|
_k = _ops.convert_to_tensor(k, _dtypes.int32)
k_input = _k
diag_index = k_input

After executing this path, related variables need to satisfy the following condition:
|\textbf{[Condition]}|
TensorShapeUtils::IsVector(diag_index.shape())
You need to conduct the following steps:
...
To help you summarize the constraint, here are the type and properties of `k'.
|\textbf{[Argument Type]}|
{'k': 'tensor'}
|\textbf{[`properties` for `k`]}|
.ndims: int, number of dimensions of tensor
.shape: [int], shape of tensor
...
|\underline{\textbf{LLM}}:|
k.ndims == 1
\end{minted}
\caption{An example prompt that instructs the LLM to infer an input constraint from the {\mycode{}if} statement in the execution path of Figure~\ref{fig:path_constraint_demonstration}. The prompt is simplified.}
\label{lst:methodology_input_condition_prompt}
\end{listing}

Although powerful, LLM can still make mistakes when inferring input constraints.
One common mistake is that LLM may infer constraints that are not solvable to the Z3 solver, which we call syntax invalid constraints.
In particular, we identify three causes of syntax invalid constraints: (1) the inferred input constraint is not targeting API inputs, (2) the inferred constraint involves undefined symbols, and (3) the inferred constraint is not grammatically correct.
To facilitate the LLM in inferring syntax valid input constraint, \toolname{} utilizes the following checks:
\begin{enumerate}[topsep=0pt, leftmargin=*]
\item The inferred input constraint should contain at least one of the API inputs (\eg, {\mycode{}k} in Listing~\ref{lst:methodology_input_condition_prompt}) and it must not involve other local variables. This check ensures that the input constraint is related to the API inputs.
\item The inferred input constraint should not include undefined symbols, which include undefined function calls, undefined variable names, or undefined input properties defined by Table~\ref{tab:api_properties}.
\item The input constraint should be formatted following the grammar of the Z3 solver. This check ensures that the input constraint can be solved by the Z3 solver.
\end{enumerate}
Violating any of these checks indicates a syntactically invalid input constraint, which is not useful for test input generation.
During the input constraint inference, \toolname{} will repeat the inference process three times until a syntactically valid input constraint is generated. 
The constraint constructed from the sanity check statement or conditional statement will be skipped by \toolname{} if no valid input constraint is generated after three attempts.

\begin{algorithm}[htbp]
\caption{Path Constraint Extraction}\label{alg:path_constraint_extraction}
\LinesNumbered{}
\KwIn{
    \mypath, the execution path whose path constraint to be extracted; 
}
\KwIn{
    \myinput, the API' inputs
}

\KwOut{\myPC, the extracted path constraint}
\SetKwProg{Def}{def}{:}{}
\Def{\AlgPCExtraction(\mypath, \myinput)}{
    \myPC{} $\leftarrow \varnothing$ \\
    \For{\Stmt in \mypath}{  
        \If{\Stmt.type is \SanityCheckStmt or \ConditionalStmt}{  
            \LocalCons $\leftarrow$ $\AlgConstraintConstruction(\Stmt)$ \\   
            \LocalVars $\leftarrow$ \LocalCons \\   
            \If{$\neg Related(\LocalVars)$}{\textbf{continue}}   
            \DF $\leftarrow$ $\AlgTraverse(\mypath, \LocalVars, \myinput)$ \\   
            \If{\AlgContainExternal(\LocalCons, \DF)}{
                iteration $\leftarrow$ 1 \\
                \While(){iteration <= 3}{   
                \InputCons $\leftarrow$ \AlgInputConstraintInference(\DF, \LocalCons, \InputVars)  // \S~\ref{subsubsec:input_constraint_inference} \\  
                \If{\AlgSyntaxValid(\InputCons)}{
                    \textbf{break};
                }
                iteration += 1\\
                }
            }
            \Else{
                \InputCons $\leftarrow$ \AlgExtractConstraint(\DF, \LocalCons, \InputVars)  // \S~\ref{subsubsec:constraint_construction}  
            }
            \If{Solvable(\InputCons)}{  
                \myPC{}.add(\InputCons)  
            }
        }
    }
    \textbf{return $\mathbf{PC}$\;}
}
\end{algorithm}

\subsubsection{Path Constraint Extraction Workflow}
Algorithm~\ref{alg:path_constraint_extraction} illustrates the detailed workflow of path constraint extraction.
When provided with an execution path, \toolname{} traverses the path, extracting conditions from sanity check statements and conditional statements (lines 3--5).
In handling loops within the path, \toolname{} follows the existing work~\cite{acetest} to unroll each loop once and ignore loop conditions.
For every extracted condition (\LocalCons), \toolname{} identifies its involved variables (\LocalVars) (line 6).
If no involved variables are related to the API inputs, the extracted condition is deemed irrelevant to the API inputs, and it will be excluded (lines 7--8).
After filtering irrelevant conditions, \toolname{} extracts the flow (\DF) from \myinput to \LocalVars{} (line 9) to construct input constraints from these conditions.
If neither \DF{} nor \LocalCons{} contain external functions, \toolname{} leverages \S~\ref{subsubsec:constraint_construction} to construct input constraints from \LocalCons.
In cases where external functions are present, \toolname{} leverages the input constraint inference (\AlgInputConstraintInference) to iteratively infer the input constraint using LLM (lines 12--17).
\toolname{} repeats the input constraint inference process until a syntactically valid input constraint is derived or the maximum number of attempts (three times) is exhausted.
Finally, the generated input constraint will be added to the path constraint if it is solvable to the Z3 solver (lines 19--20).

\subsection{Step III. Test Input Generation and Bug Detection}\label{subsec:method_input_generation}
After extracting path constraints from both the API under test and its counterpart, \toolname{} utilizes them to generate test inputs iteratively.
In each iteration, \toolname{} selects one path constraint to generate a test input for differential testing.

\textbf{Path Constraint Selection}. Since an API can have multiple path constraints, a path constraint selection process is necessary. 
Following previous works~\cite{muffin,comet}, \toolname{} adopts a basic Roulette Wheel Selection~\cite{back1996evolutionary} algorithm with an intention that path constraints rarely selected should have a higher chance to be selected.
Specifically, for each path constraint, \toolname{} tracks the number of iterations that it has been selected, denoted as $c$.
Then \toolname{} defines the fitness score of each path constraint as $s = \frac{1}{c+1}$.
The selection probabilities for each path constraint are defined as follows:

\begin{equation}
    p = \frac{s_i}{\sum_{i=1}^{N}s_i}
    \label{eq:pc_selection}
\end{equation}

Here, $N$ presents the total number of path constraints. 
In each iteration, \toolname{} selects a path constraint based on its probability defined in Equation~\ref{eq:pc_selection}, then \toolname{} leverages the selected path constraint for testing and bug detection.

\textbf{Testing and Bug Detection}. 
Inspired by TensorScope~\cite{tensorscope}, before using the selected path constraint for test input generation, \toolname{} augments it with additional input constraints, including duplicate-input constraints, natural constraints, property-value constraints, and error-feedback constraints.

\begin{enumerate}[topsep=0pt, leftmargin=*]
\item \textbf{Duplicate-input constraints}. These constraints prevent the repetition of previously generated test inputs. 
\toolname{} gathers property values from past inputs and constructs these constraints to avoid generating test inputs with the same property values.
\item \textbf{Natural-constraints}. Natural constraints define the value space of input argument's properties, ensuring property values conform to specified ranges. For instance, tensor's {\mycode{}ndims} must fall within [0,5].
\item \textbf{Error-feedback constraints}. 
\toolname{} uses predefined patterns to extract error-feedback constraints from error messages in previous iterations, preventing similar invalid inputs from being generated. 
For instance, consider the error message: \textit{Dimension out of range (expected to be in range of [\#1, \#2], but got \#3)}, suggesting that a tensor input's {\mycode{}ndims} should fall within [\#1, \#2], but the actual value is \#3.
Regarding this error message, \toolname{} adds a constraint restricting the {\mycode{}ndims} of the tensor input, originally \#3, to be in the range of [\#1, \#2].
\item \textbf{Property-value constraints}. \toolname{} extracts property-value constraints from validation inputs generated in \cref{subsubsec:validation_input_generation}.
Since validation inputs are constructed via exploring different property values and including those valid ones, if an argument holds the same property value among all validation inputs, this argument property with a different value may be invalid.
Based on this assumption, property-value constraints restrict an argument's property to be a fixed value if it consistently holds the same value among all validation inputs.
For instance, if a tensor argument {\mycode{}x}'s {\mycode{}ndims} is always 1 in validation inputs, \toolname{} adds the constraint {\mycode{}x.ndims==1} to fix its {\mycode{}ndims} to 1.
While facilitating test input validity, this constraint may limit the input diversity since certain property values are fixed.
To allow exploring different values for these properties, \toolname{} assigns a low utilization probability (\ie, 0.3) to this constraint during the test input generation, ensuring that these property values are not always fixed.
\end{enumerate}

\begin{algorithm}[thbp]
    \caption{Path Constraint Selection And Differential Testing}\label{alg:test_input_generation}
    \LinesNumbered
    \KwIn{
        $\myPC{}s$: path constraints; \undertest: API under test; \counterpart: \undertest{}'s counterpart; N: max iterations
    }
    \KwOut{Test Input}
    \SetKwProg{Def}{def}{:}{}
    \Def{\AlgTestInputGeneration(\myPC{}s, \undertest{}, \counterpart{}, N)}{
        iteration $\gets 0$ \\
        \While(){iteration $< N$}{
            $\myPC{}\gets \PCSelector.select(\myPC{}s)$ \\
            \myPC{}s.remove(\myPC{}) \\  
            \myPC{}.add(\NaturalCons) \\   
            \textbf{with probability} $p = 0.3$: \myPC{}.add(\PropertyCons)  \\  
            $solution \gets Z3(\myPC)$ \\
            $\TestInput \gets solution$ \\   
            $DifferentialTesting$(\TestInput,\undertest{},\counterpart{})  \\
            \If{$crash(\TestInput) \land error\_msg \in InvalidErrorSet$}{
                $\ErrorCons \gets error\_msg$  \\
                \myPC{}.add(\ErrorCons)   
            }
            $\DuplicateCons \gets \TestInput$ \\
            \myPC{}.add(\DuplicateCons) \\   
            \myPC{}s.add(\myPC{}) \\   
            iteration $\gets$ iteration + 1  \\
        }
    }
\end{algorithm}

Algorithm~\ref{alg:test_input_generation} illustrates the detailed test input generation process.
Given path constraints \myPC{}s extracted from an API \undertest{} and its counterpart \counterpart, \toolname{} iteratively select one path constraint (line 4) from \myPC{}s for test input generation.
In each iteration, the selected path constraint will be enriched with natural constraints (\NaturalCons, line 6). 
The property-value constraints (\PropertyCons) with a low utilization probability ($p=0.3$) may be further added to \myPC{} (line 7).
\toolname{} leverages a Z3 solver (\ie, z3py~\cite{z3py}) to solve a solution satisfying \myPC{} (line 8).
Based on this solution, a fuzzing driver is used to instantiate the concrete input values for differential testing and bug detection (lines 9--10).
At the end of each iteration, \toolname{} updates the error-constraint (\ErrorCons) and duplicate-constraint (\DuplicateCons) identified in this iteration to \myPC{} (lines 11--15).
The updated path constraint will be added back to \myPC{}s for future test input generation.

Inspired by the existing work~\cite{numerical_discrepancies_issta} demonstrating that special values (\ie, {\mycode{}Inf} and {\mycode{}NaN}) can expose output inconsistency in scientific libraries, we also randomly generate some special values (with probability $p=0.05$) when instantiating tensor values from a solution (line 9 in the Algorithm~\ref{alg:test_input_generation}).

During the bug detection, \toolname{} targets the following three types of bugs.
\begin{itemize}[leftmargin=*,topsep=0pt]
    \item \textbf{Incorrect result bug}. This type of bug refers to incorrect computation results of DL library APIs. 
    The incorrect result bug is detected by measuring the output difference between the API under test and this API's counterpart. 
    Following existing works~\cite{titanfuzz,fuzzgpt}, we use a tolerance threshold (\ie, 0.1) to check if output differences are significant.
    \item \textbf{Incorrectly-rejected bug}. Incorrectly-rejected bugs refer to unexpected runtime exceptions of DL library APIs.
    This bug is captured by checking the execution status inconsistency between the DL library API and its counterpart.
    For each execution status inconsistency, we remove clear syntax errors (\eg, {\mycode{}SyntaxError}) and invalid argument errors such as {\mycode{}TypeError} and {\mycode{}RuntimeError}~\cite{freelunch}.
    \item \textbf{Crash bug}. Crash bugs refer to system crashes, including aborts, segmentation faults, floating point exception raised, and `{\mycode{}INTERNAL\_ASSERT\_FAILED}'~\cite{fuzzgpt, tensorscope}. 
\end{itemize}

\subsection{Implementation}
We use the \llm{}, one of the most popularly large language models, for counterpart synthesis and path constraint extraction.
Following TitanFuzz~\cite{titanfuzz}, we set the temperature to 0.4. 
To better elicit LLM's power, we also employ two commonly used prompting strategies when instructing LLM for validation input generation and input constraint inference.
During the LLM prompting for validation input generation, inspired by a popular self-debugging strategy~\cite{llm_self_debug}, we prompt the \llm{} with the error message for self-debugging once if \llm{} generates an invalid input.
Listing~\ref{lst:input_repairs} exemplifies this self-debugging workflow, while the LLM initially generates an invalid input, it successfully generates a valid one after we provide the error message. 
As for input constraint inference, since the correctness of inferred input constraint cannot be verified through dynamic execution, the self-debugging strategy is not applicable.
Instead, we follow established approaches~\cite{llift,weng2022large} to prompt the LLM with a self-validation message (refer to Listing~\ref{lst:self_validation_constraint_prompt}). 
This prompt encourages the LLM to review its previous steps, aiming for self-correction by recognizing and rectifying potential mistakes.
Finally, inspired by a recent work~\cite{llift}, we augment our prompt with additional instructions (one for counterpart synthesis and six for input constraint inference), intending to guide the LLM for better results.
For example, we restrict the LLM to only use PyTorch APIs when synthesizing counterparts for TensorFlow APIs, and we restrict the LLM to only synthesize constraints for given input arguments' properties.
We release the detailed prompt instructions in our artifact~\cite{DLLens}.

\begin{listing}[htbp]
\begin{minted}[
    baselinestretch=1.0,
    fontsize=\scriptsize,
    xleftmargin=0.5ex,
    bgcolor=bg,
    breaklines=true,
    escapeinside=||,
]{text}
|\underline{\textbf{Prompt}}|:
Task 1: Import TensorFlow 2.10.0
Task 2: Generate valid parameter. The name of parameter variables should be: `image', ...
Task 3: Call the function: tf.image.pad_to_bounding_box(image, ...)
|\underline{\textbf{LLM}}:|
image=tf.constant([[1,2],[3,4]],dtype=tf.float32) |\textbf{\textcolor{bgRed}{(the original input is invalid!)}}|
|\underline{\textbf{Follow-up Prompt}}|:
[Error Message]
'image' (shape(2,2)) must have 3 or 4 dimensions.
|\underline{\textbf{LLM}}|: Here is the repaired input
image=tf.constant([[[1,2],[3,4]]],dtype=tf.float32) |\textbf{\textcolor{bgGreen}{(the self-corrected input is valid.)}}|
\end{minted}
\caption{An example prompt that instructs the LLM to self-debug its previously generated invalid input.}
\label{lst:input_repairs}
\end{listing}

\begin{listing}[htbp]
\begin{minted}[
    baselinestretch=1.0,
    fontsize=\scriptsize,
    xleftmargin=0.5ex,
    bgcolor=bg,
    breaklines=true,
    escapeinside=||
]{text}
Review the assistant analysis above carefully; consider the following:
- The generated constraint should be a boolean expression that only includes the symbols [INPUT_ARGUMENTS] and their attributes.
- If the constraint includes attributes or arguments that are not explicitly required by the given condition, you should not include them in the final answer.
- If the argument and argument name is correct.
Thinking step by step, conclude a correct and comprehensive answer
\end{minted}
\caption{The prompt that instructs the LLM to conduct self-validation on its inferred input constraint.}
\label{lst:self_validation_constraint_prompt}
\end{listing}

\begin{table}[htbp]
\caption{Constraint Construction Rules for Sanity Check Statements.}
\renewcommand{\arraystretch}{0.6}  
\resizebox{0.6\linewidth}{!}{
\begin{tabular}{l|l}
\toprule
Sanity Check Statement &   Extracted Condition\\ \midrule
assert({\mycode{}arg1}) &   `{\mycode{}arg1}'  \\ \midrule  
TORCH\_CHECK*(\mycode{}arg1) &   `{\mycode{}arg1}'  \\ \midrule  
OP\_REQUIRES*({\mycode{}arg1}, {\mycode{}arg2}, {\mycode{}arg3}) & `{\mycode{}arg2}'  \\ \midrule  
args\_to\_matching\_eager([{\mycode{}arg1}], {\mycode{}arg2}, [{\mycode{}arg3}]) & `{\mycode{}arg1}.dtypes in [{\mycode{}arg3}]' \\ \bottomrule  
\end{tabular}
}
\label{tab:sanity_check_constraints}
\end{table}

To collect the source code of DL library APIs, we use PyCG~\cite{pycg} and Joern~\cite{joern}/tree-sitter~\cite{treesitter} to construct call graphs for Python and C++, respectively.
Starting from the entry point of each API, we conduct a breadth-first search in the constructed call graph, collecting the source code of all functions within a depth of five levels.
Note that TensorFlow's and PyTorch's front-end APIs are implemented in Python, and their core logic functions are implemented in C++. 
To establish the call relationship between the Python and C++ code, we observe that TensorFlow employs the Python interface \texttt{\_execute.execute} to access C++ operators, and PyTorch utilizes \texttt{native\_functions.yaml} to link the Python API with its C++ implementations. 
Leveraging these characteristics, we develop a program analysis tool to bind function calls between Python and C++.

To extract path constraints from collected source code, we use the Python standard AST package~\cite{python_ast} and tree-sitter~\cite{treesitter} to parse code into syntax trees.
Subsequently, our static analysis tool is used to extract path constraints for each execution path within the control flow graphs built on these syntax trees.
When constructing constraints from sanity check statements, we manually implemented four rules for four types of validation functions popularly used in TensorFlow and PyTorch. 
We summarize these rules in Table~\ref{tab:sanity_check_constraints}.

\section{Evaluation}

We evaluate \toolname{}'s performance with four research questions (RQs).

\begin{itemize}[leftmargin=*,topsep=0pt]
    \item \textbf{RQ1: How effective is \toolname{} in synthesizing counterparts?} 
    Understanding how many APIs \toolname{} can synthesize counterparts for is essential since differential testing may be inapplicable for an API if the API's counterpart is unavailable.
    Besides the number of APIs supported by \toolname{}, we also evaluate the quality of our synthesized counterparts.
    \item \textbf{RQ2: How effective is \toolname{} in path constraint extraction?}
    The extracted path constraints are important in guiding test input generation for diverse execution path exploration.
    We use multiple metrics to evaluate the extraction effectiveness of \toolname{}, including the number of property constraints extracted for API inputs, unique path constraints extracted, and input constraints extracted per path.
    \item \textbf{RQ3: Are \toolname{} more effective in testing than existing works on a set of randomly sample APIs?}
    We compare \toolname{} with existing techniques regarding the branch coverage and the number of bugs detected on 200 randomly sampled DL library APIs. 
    \item \textbf{RQ4: Can \toolname{} detect new real bugs?}
    Following prior works on DL library testing~\cite{freelunch,docter,deeprel,acetest,tensorscope,titanfuzz,fuzzgpt}, we report the number of real bugs detected to demonstrate the usefulness of \toolname{}.

\end{itemize}

\noindent \textbf{Environment.}  The experiments are conducted using a 32-core server with four 3090Ti GPUs\@.

\subsection{Baseline and Subject Selection}

We consider six state-of-the-art techniques, TitanFuzz~\cite{titanfuzz}, ACETest~\cite{acetest}, TensorScope~\cite{tensorscope}, FreeFuzz~\cite{freelunch}, DeepREL~\cite{deeprel}, and DocTer~\cite{docter} in our evaluation. 
Since neither executables nor source codes of FuzzGPT~\cite{fuzzgpt} are available, we do not consider it in 
evaluation.\footnote{Although the source code of TensorScope is available, we try our best but fail to execute their constraint extraction and input generation modules.
Thus, we do not include TensorScope in RQ2 and RQ3.}

In RQ1, we compare \toolname{} with TensorScope~\cite{tensorscope}, which, to our best knowledge, is the only existing tool that collects API counterparts across DL libraries.
TensorScope collects counterparts of TensorFlow and PyTorch APIs via parsing the conversion rules in model conversion libraries.
Given that these rules are defined by library developers, who likely have high expertise in DL libraries, we consider all counterparts extracted from these rules to be valid.

In RQ2, we compare \toolname{} with DocTer~\cite{docter}, the state-of-the-art work mining input constraints from DL library API documentation. 
Although ACETest~\cite{acetest} also considers constraints in their methodology, it discloses only the extracted constraints (in SMT formula format)~\cite{acetest_repo} without providing the source code to extract them, making it difficult to compare the APIs beyond the disclosed ones. 
Additionally, we observe that the formulas released by ACETest can contain duplicated input constraints.
This is possibly due to the adoption of a different criterion for path constraint counting.
Specifically, ACETest counts all path constraints extracted from each execution path, while we count only those with a unique set of input constraints.
Directly comparing \toolname{} with the results reported by ACETest without reproduction and deduplication can be unfair. Therefore, we omit the path constraint comparison with ACETest in RQ2. Instead, we compare its overall testing performance with \toolname{} in RQ3.

In RQ3, we compare \toolname{} with all the five state-of-the-art DL library testing techniques: TitanFuzz~\cite{titanfuzz}, ACETest~\cite{acetest}, DocTer~\cite{docter}, FreeFuzz~\cite{freelunch}, and DeepREL~\cite{deeprel} on a 200 randomly sampled APIs. 

We evaluate \toolname{} on the APIs from the two most popular DL libraries, PyTorch (v2.1.0)~\cite{pytorch} and TensorFlow (v2.10.0)~\cite{tensorflow}.
We exclude those APIs that
(1) have an incomplete signature, or
(2) belong to a special or experimental package (\eg, {\mycode{}torch.utils} or {\mycode{}tf.experimental}). 
In total, we select 1,862 and 1,121 APIs from TensorFlow and PyTorch, respectively, as our experiment subjects.
We release the information of the selected APIs and selection criteria in our artifact~\cite{DLLens}.

\subsection{RQ1: Effectiveness of Counterpart Synthesis}

\subsubsection{Metrics}
We start with evaluating the capability of \toolname{} in synthesizing counterparts. 
Specifically, we use the following two metrics.
\begin{itemize}[leftmargin=*,topsep=0pt]
    \item \textbf{Number of API Counterparts}. 
    We compare \toolname{} with TensorScope by measuring the number of DL library APIs whose counterparts can be successfully synthesized by each tool. 
    \item \textbf{Diversity of Validation Inputs}. We measure the diversity of validation inputs on which \toolname{}'s synthesized counterparts are deemed valid (\ie, satisfying the Equation~\ref{eq:counterpart_definition}). A higher diversity suggests a more comprehensive evaluation of the synthesized counterparts.
    To measure the diversity of validation inputs for each counterpart, we follow COMET~\cite{comet} and report the number of unique argument property values regarding the eight properties defined in Table~\ref{tab:api_properties}.
    Since TensorScope does not use validation inputs to validate their counterparts, we do not compare \toolname{} with it on this metric.
\end{itemize}

\subsubsection{Setup}
\noindent To mitigate the impact of hallucination issues in LLM~\cite{Liu2024ExploringAE}, we iteratively execute \toolname{} to synthesize counterparts for each API five rounds until a validated counterpart is synthesized (\ie, deemed valid on validation inputs).
Table~\ref{tab:eval_counterpart} outlines the number of APIs for which \toolname{} successfully synthesizes counterparts in each round.
In total, \toolname{} successfully synthesizes valid counterparts for 1,401 APIs, with 689 from TensorFlow and 712 from PyTorch, accounting for 37.00\% and 63.51\% of the total APIs evaluated, respectively.
The five-round counterpart synthesis process took 6.42 hours using 20 processes. 
The total costs for the RQ1 experiment were approximately \$20 in \llm.
Given that this counterpart synthesis represents a one-time analysis per library, we consider the time and financial costs acceptable.

\subsubsection{Results}
\begin{table}[htbp]
\renewcommand{\arraystretch}{0.7}
\caption{The number of API counterparts found by \toolname{} in each round.}\label{tab:eval_counterpart}
\centering
\resizebox{0.8\linewidth}{!}{
\begin{tabular}{@{}l|rrrrr|l@{}}  
\toprule
Library Name & Round 1 & Round 2 & Round 3 & Round 4 & Round 5 & Total \\ \midrule
TensorFlow   & 604     & + 37      & + 17      & + 14      & + 17      & 689  \\
PyTorch      & 608     & + 47      & + 34      & + 12      & + 11      & 712   \\ \bottomrule
\end{tabular} %
}
\end{table}

\noindent \textbf{Comparison With TensorScope}. We compare \toolname{} with TensorScope~\cite{tensorscope}, the only work collecting API counterparts across DL libraries.
To implement TensorScope, we adhere to its methodology and collect counterparts of TensorFlow and PyTorch APIs via parsing the conversion rules in model conversion libraries.

\begin{figure}[t!]
    \centering
    \includegraphics[width=0.6\linewidth]{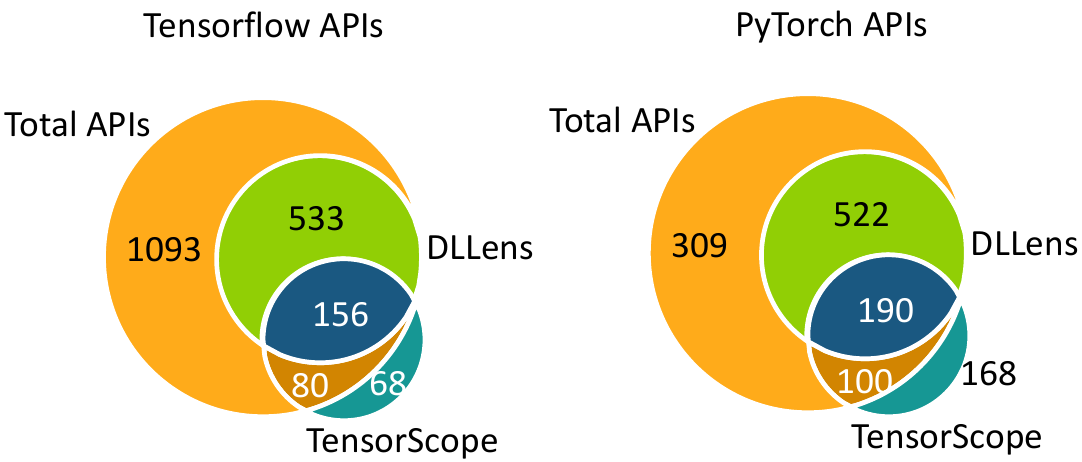}
    \caption{Comparison on the number of API counterparts found by \toolname{} and TensorScope.}\label{fig:api_coverage}
    \Description{Comparison on API Counterpart Coverage}
\end{figure}

Figure~\ref{fig:api_coverage} compares the number of API counterparts found by \toolname{} and TensorScope.
Overall, \toolname{} significantly outperforms TensorScope by collecting 126.64\% (689 v.s. 304) more counterparts for TensorFlow APIs and 55.46\% more for PyTorch APIs (712 v.s. 458).
In particular, \toolname{} collects counterparts for 1,055 (533+522) APIs where TensorScope fails.
This improvement is crucial as it makes differential testing applicable to a broader range of DL library APIs, thus increasing the chance of detecting more bugs.
Considering the case of the buggy API {\mycode{}tf.math.is\_non\_decreasing} in Listing~\ref{lst:motivating_bug}. 
TensorScope does not collect any counterpart for this buggy API, making differential testing inapplicable to detect its output incorrectness.

Note that TensorScope found 68 and 168 counterparts for TensorFlow and PyTorch APIs not included in our API records.
These additional APIs found by TensorScope, based on our observation, are primarily related to specialized packages (\eg, {\mycode{}torch.utils}) or unique uses (\eg, {\mycode{}torch.float64} for data type representation).
We do not include these APIs in our API records since our emphasis is on APIs developed for DL algorithms, which may be commonly used by DL library users. 
Even taking into account these APIs, the number of covered APIs of TensorScope is still less than our result.

\begin{figure}[t!]
    \centering
    \includegraphics[width=1.0\linewidth]{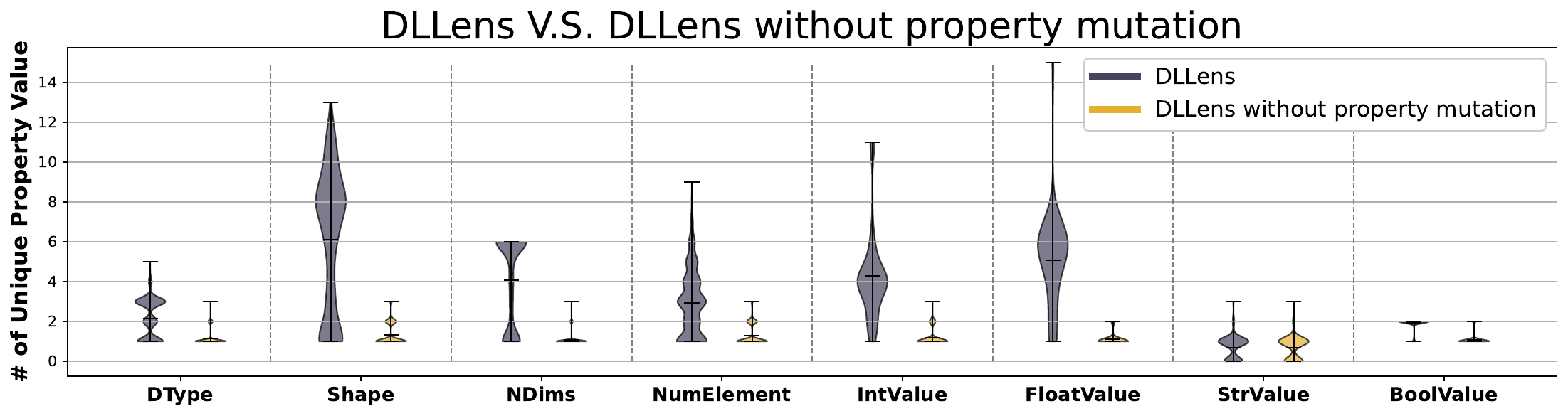}
    \caption{Comparison on the distribution of unique property values.
    }\label{fig:dist_validation_inputs}
    \Description{Distribution of Validation Inputs}
\end{figure}

\noindent \textbf{Effectiveness of Property Mutation}. 
We further evaluate the diversity of our validation inputs and the effectiveness of property mutation.
Property mutation aims to diversify inputs generated from LLM prompting to enable the validation of synthesized counterparts across a wider range of inputs.
Figure~\ref{fig:dist_validation_inputs} compares the diversity of validation inputs generated by \toolname{} with and without property mutation.
When relying solely on LLM prompting (\ie, \toolname{} without property mutation), \toolname{} only generates a maximum of three unique values for all argument properties.
In particular, LLM prompting tends to generate the same data type, shape, and number of dimensions for a tensor arguments.
In contrast, property mutation significantly increases in the number of unique property values for each argument.
For instance, in the case of tensor arguments, property mutation increases the number of unique values for properties such as {\mycode{}dtype} (from 1.13 to 2.15), {\mycode{}shape} (from 1.33 to 6.12), {\mycode{}ndims} (from 1.08 to 4.07), and {\mycode{}num\_element} (from 1.30 to 2.94).

\noindent \textbf{Validity of Counterparts}. We conduct a manual verification to understand the validity of counterparts synthesized by \toolname{}.
To do so, a random selection of 50 counterparts from both TensorFlow and PyTorch in Table~\ref{tab:eval_counterpart} is sampled for manual verification.
For each synthesized counterpart, the manual analysis aims to identify if its functionality is consistent with its target API by checking both the documentation of the target API and the counterpart's implementation.
The manual analysis is conducted independently by two authors of this paper, and the final results reach a unanimous agreement among both authors.
The manual verification reveals that most counterparts (\ie, 81 out of 100) implement the same functionalities as their target APIs.
However, discrepancies are observed in the functionalities of the remaining 19 counterparts under specific input conditions.
We categorize these 19 counterparts as partially valid, as their functionalities are still valid for some input conditions.

Listing~\ref{lst:examplecounterpart_partially_valid} shows a partially valid counterpart for API {\mycode{}tf.raw\_ops.Bincount}.
Both the API and its counterpart aim to count the occurrences of each value within an integer array.\footnote{\url{https://www.tensorflow.org/api_docs/python/tf/raw_ops/Bincount}.}
This counterpart is verified as partially valid since it aligns with the API only when {\mycode{}size >= max(arr)}, but their functionality diverges when {\mycode{}size < max(arr)}.  
For instance, when {\mycode{}size} is smaller than {\mycode{}max(arr)} (\eg, {\mycode{}size=1,arr=[1,2,2,3]}), {\mycode{}tf.raw\_ops.Bincount} only outputs occurrences of values smaller than {\mycode{}size} (\eg, outputs {\mycode{}[0.]}).  
In contrast, the counterpart disregards the {\mycode{}size} constraint and reports occurrences for all values (\eg, outputs {\mycode{}[0,1,2,1]}).
Since validation inputs generated by \toolname{} omitted the condition {\mycode{}size < max(arr)}, this discrepancy was not exposed during our counterpart validation process.  

\begin{mdframed}[style=MyFrame]
\textbf{Summary of RQ1}: 
\toolname{} successfully synthesizes counterparts for 126.64\% (689 v.s. 304) more TensorFlow APIs and 55.46\% (712 v.s. 458) more PyTorch APIs as compared with the state-of-the-art approach (TensorScope~\cite{tensorscope}).
Additionally, our proposed property mutation technique can improve the diversity of validation inputs, enabling a more comprehensive validation on synthesized counterparts.
Our manual analysis on 100 randomly sampled counterparts reveals that most (81) counterparts' functionalities align with their target APIs, with the rest deemed partially valid.
Due to the small portion of partially valid counterparts, we apply manual filtering to address false positives introduced by them (see \S~\ref{sec:discussions}).
\end{mdframed}

\begin{listing}
\begin{minted}[
    baselinestretch=1.0,
    fontsize=\scriptsize,
    xleftmargin=0.5ex,
    bgcolor=bg,
    breaklines=true,
    escapeinside=||,
]{text}
|\underline{\textbf{TensorFlow API}}|
tf.raw_ops.Bincount(arr,size,weights)
|\underline{\textbf{Counterpart}}|
def pytorch_call(arr, size, weights):
    weights = weights.view(-1)
    counts = torch.bincount(arr, weights=weights, minlength=size.item())
    return counts
|\underline{\textbf{Example Input on Which The Counterpart is Invalid}}|
arr = [1,2,2,3]
size = 1
weights = [1,1,1,1]
|\underline{\textbf{API's Output V.S. Counterpart's Output}}|
[0.] v.s. [0.,1.,2.,1.]
\end{minted}
\caption{A partially valid counterpart of API {\mycode{}tf.raw\_ops.Bincount}. The counterpart has a different functionality when {\mycode{}size < max(arr)}.}
\label{lst:examplecounterpart_partially_valid}
\end{listing}

\subsection{RQ2: Effectiveness of Path Constraint Extraction}\label{sec:rq2}

\subsubsection{Metrics}
We use the following three metrics to evaluate the constraint extraction capability of \toolname{}.
\begin{itemize}[leftmargin=*,topsep=0pt]
    \item \textbf{Number of Property Constraints}. To compare with DocTer~\cite{docter}, we report the number of property constraints extracted for each API\@. The property constraint is commonly used by existing works~\cite{tensorscope,docter}, which refers to the total number of input argument properties considered by extracted constraints. For each API input argument, we follow DocTer and consider all valid options for a property as one property constraint.
    \item \textbf{Number of Unique Path Constraints}. We report the average number of unique path constraints extracted from each DL library API's source code\@. 
    Each unique path constraint contains a unique set of constraints on API's inputs.
    For each API, its path constraints are formed by combining path constraints extracted from both its own implementation and its counterpart's implementation.
    Since DocTer does not extract path constraints, we compare \toolname{} with its baseline version (\toolnamenonllm{}) on this metric.
    The baseline version extracts path constraints without using LLM-based input constraint inference.
    \item \textbf{Number of Input Constraints Per Path}. For each path constraint, we also report the number of input constraints inside it. We define each valid constraint extracted from a sanity check or conditional statement as an individual input constraint. The more input constraints included in a path constraint, the more complete the path constraint is likely to be.
    We compare \toolname{} with \toolnamenonllm{} on this metric.
\end{itemize}

\subsubsection{Setup}
\noindent We apply \toolname{} to extract path constraints from all APIs (\ie, 689 TensorFlow APIs and 712 PyTorch APIs) and their counterparts synthesized in RQ1.
To assess the effectiveness of the LLM-based input constraint inference, we conducted a comparative analysis between \toolname{} and its baseline version without this input constraint inference (\toolnamenonllm).
In total, the extraction process with \toolnamenonllm{} took around 1.00 hours, utilizing 20 processes.
As for \toolname, which uses the \llm{} API for input constraint inference, it took a longer time (3.34 hours) with 20 processes to extract path constraints. 
The total cost of utilizing the \llm{} API during the RQ2 experiment is estimated to be around \$15.
Similar to the counterpart synthesis evaluation in RQ1, our constraint extraction is also a one-time analysis per library, thus we consider the time and financial costs acceptable.

\subsubsection{Results}
\textbf{Number of Property Constraints}. We first compare \toolname{} with constraints released by DocTer~\cite{docter}.
We utilize the number of property constraints as the metric, which is adopted by DocTer, to evaluate the constraint extraction performance.
Among all 1,401 APIs for which \toolname{} synthesized counterparts in RQ1, we notice that DocTer only covers the constraints for 498 APIs (345 TensorFlow and 153 PyTorch APIs)~\cite{docter_repo}.
For a fair comparison, we compare the property constraint extraction performance with DocTer on these 498 APIs.
Table~\ref{tab:eval_property_constraint} shows the average number of property constraints extracted for each API and the total number of property constraints extracted by each tool\@.
For each API, \toolname{} averagely extracts 6.30 property constraints while DocTer extracts 4.34.
In total, \toolname{} outperforms DocTer by extracting 45.07\% (3,135 v.s. 2,161) more property constraints.
Digging deeper, we further analyze the constraint extraction performance for each property, as shown in Table~\ref{tab:eval_property_constraint}. 
Compared with DocTer, we observe that \toolname{} can extract more shape (2.31 vs. 0.81), value (0.46 vs. 0.17), and structure (2.28 vs. 1.87, `STR.' in Table~\ref{tab:eval_property_constraint}) constraints per API\@.
As for data type (`DType' in Table~\ref{tab:eval_property_constraint}) constraints, averagely \toolname{} extracts 1.24 constraints per API, whereas DocTer surpasses \toolname{} by extracting 1.49 per API\@. 
The superior performance of DocTer in extracting data type constraints could be due to the common practice of documenting constraints related to data types, making DocTer more effective in this aspect. 
In contrast, constraints on shape, value, and structure are less frequently documented, limiting DocTer's performance in these properties.
\begin{table}[t!]
\caption{Comparison on the number of property constraints extracted from the concerned 498 APIs.}\label{tab:eval_property_constraint}
\newcommand{\twocol}[1]{\multicolumn{2}{c}{#1}}
\newcommand{\tworow}[1]{\multirow{2}{*}{#1}}
\newcommand{\fourcol}[1]{\multicolumn{4}{c}{#1}}
\centering
\renewcommand{\arraystretch}{1.0}
\resizebox{0.7\linewidth}{!}{

\begin{threeparttable}
\begin{tabular}{l|r|r|rrrr@{}}
\toprule
Tool          & Total Constraints & Average Constraints  &  DType        & Shape         & Value         & Struc.  \\ \midrule
\toolname{}   & \textbf{3135}     &  \textbf{6.30}       & 1.24          & \textbf{2.31} & \textbf{0.46} & \textbf{2.28} \\ \midrule
DocTer        & 2161              &  4.34                & \textbf{1.49} & 0.81          & 0.17          & 1.87 \\\bottomrule
\end{tabular}
\begin{tablenotes}
\item [1] `DType' is short for `Data Type'; `Struc.' is short for `Structure'.
\end{tablenotes}
\end{threeparttable}
}
\end{table}

\textbf{Number of Unique Path Constraints}. We further evaluate the number of unique path constraints extracted by \toolname{}, as is shown in Table~\ref{tab:eval_path_constraint} (rows 2 and 4).
On average, \toolname{} extracts 18.14 path constraints for each TensorFlow API and 38.76 for each PyTorch API\@, including path constraints extracted from their counterparts.
In comparison to \toolnamenonllm{} (rows 3 and 5 in Table~\ref{tab:eval_path_constraint}), incorporating LLM-based input constraint inference resulted in a significant improvement in the number of extracted path constraints (\ie, 18.14 v.s. 2.79 for TensorFlow APIs and 38.76 v.s. 2.75 for PyTorch APIs).
The improvement can be attributed to the effectiveness of input constraint inference when handling external functions, which static analysis does not address due to the lack of their source code.

\textbf{Number of Input Constraints Per Path}. We also report the number of input constraints included in each extracted path constraint.
On average, each path constraint in TensorFlow and PyTorch APIs includes 3.18 and 3.91 input constraints, respectively.
This indicates that each path constraint captures the input constraints required by around three sanity checks or conditional statements, on average.
In comparison, \toolname{} without the input constraint inference (\ie, \toolnamenonllm) can only extract 1.78 and 1.57 input constraints per path for TensorFlow and PyTorch APIs.
This performance drop implies the necessity of input constraint inference to analyze those external functions.
Inside a path constraint, the number of input constraints plays a critical role in evaluating the quality of the path constraint. 
A higher number of input constraints could narrow down the search space for test input generation, thereby improving the effectiveness of generating test inputs to reach the target path.

\begin{table}[htbp]
\caption{Comparison on the number of unique path constraint (P.C.) extracted by \toolname{} and \toolnamenonllm.}\label{tab:eval_path_constraint}
\resizebox{0.6\linewidth}{!}{
\centering
\begin{tabular}{l|l|r|r}
\toprule
Library                     & Tool              & \# P.C./API & \# Cons./P.C. \\ \cmidrule{1-4}
\multirow{2}{*}{TensorFlow} & \toolname{}       & \textbf{18.14} & \textbf{3.18} \\ \cmidrule{2-4}
                            & \toolnamenonllm{} & 2.79 & 1.78 \\ \cmidrule{1-4}
\multirow{2}{*}{PyTorch}    & \toolname{}       & \textbf{38.76} & \textbf{3.91} \\ \cmidrule{2-4}
                            & \toolnamenonllm{} & 2.75 & 1.57 \\\bottomrule
\end{tabular}
}
\end{table}

\textbf{Correctness and Usefulness of Inferred Input Constraints}.
We further evaluate constraints inferred by our LLM-based input constraint inference.
To evaluate the correctness of these inferred constraints, we conduct a manual verification on 50 randomly sampled APIs (25 APIs for both TensorFlow and PyTorch), with the results shown in Table~\ref{tab:manual_llm_constraint}.
Among these 50 APIs, \toolname{} could leverage \llm{} to infer 694 (223 for TensorFlow and 471 for PyTorch) input constraints from related sanity checks or conditional statements.
Notably, all these statements include external functions, making \toolnamenonllm{} fail to extract any input constraints from them.
As a result, \toolnamenonllm{} could only extract 79 input constraints from these 50 APIs.
Regarding the accuracy of inferred constraints, our manual validation confirmed that 348 out of 694 (50.14\%) inferred constraints are correct. 
While not all inferences are correct, the LLM-based input constraint inference significantly increases the number of accurate input constraints compared to \toolnamenonllm{}.
This enhancement can notably reduce the search space during test input generation when exploring execution paths within these APIs.

\begin{table}[htbp]
\centering
\caption{The manual analysis results on inferred input constraints.}\label{tab:manual_llm_constraint}
\resizebox{0.6\linewidth}{!}{
\begin{threeparttable}
\begin{tabular}{l|r|r|r}
\toprule
Library    & \# total inferred & \# inferred / API   & \# correctly inferred / API \\ \midrule
TensorFlow & 223 & 8.92   & 4.56    \\ \midrule
PyTorch    & 471 & 18.84   & 9.36   \\ \bottomrule
\end{tabular}
\begin{tablenotes}
\item [1] \# total inferred: the total number of inferred input constraints; \# inferred / API: the average number of inferred input constraints per API; \# correctly inferred / API: the average number of correctly inferred input constraints per API.
\end{tablenotes}
\end{threeparttable}
}
\end{table}

\begin{figure}[t!]
\centering
\includegraphics[width=0.9\linewidth]{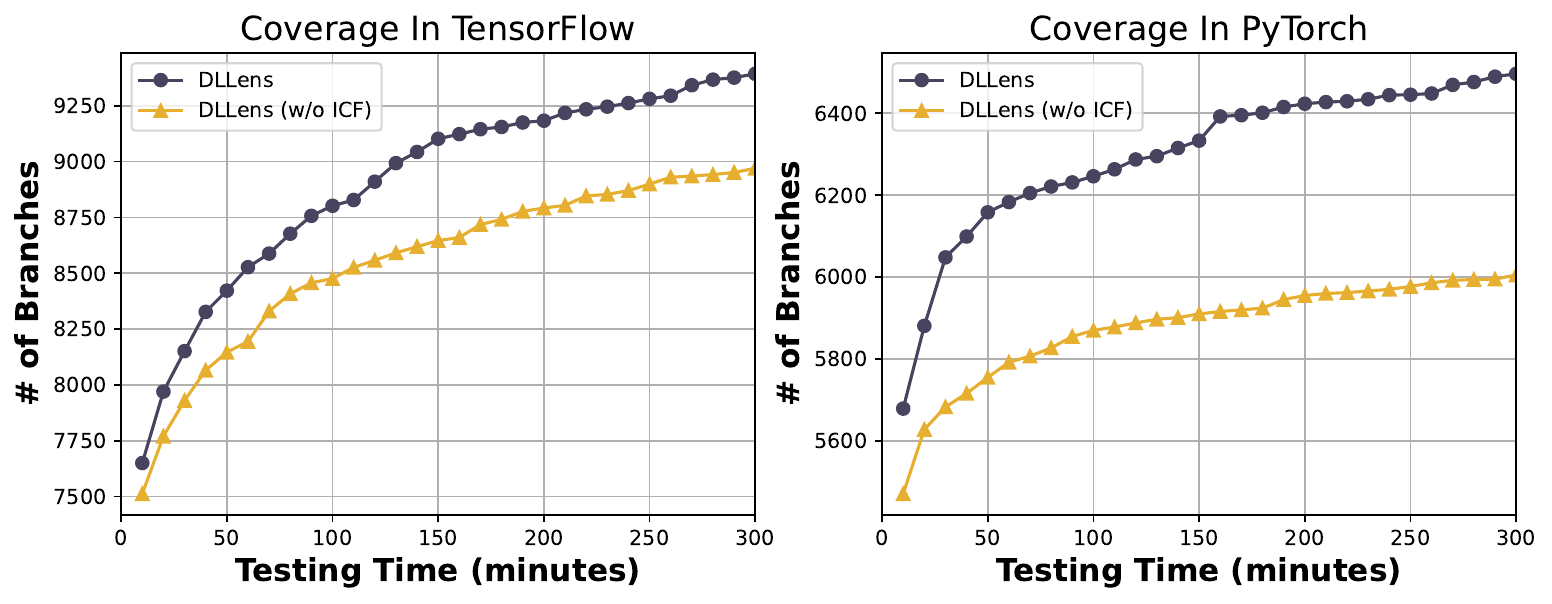}
\caption{Comparison on the code coverage achieved between \toolname{} and \toolnamenonllm{}.}\label{fig:ablation_coverage_trend}
\Description{\toolname{} v.s. \toolnamenonllm{} on Code Coverage}
\end{figure}

To demonstrate the usefulness of these inferred constraints, we conduct an ablation study by comparing the branch coverage achieved by \toolname{} and \toolnamenonllm{}.
The evaluation is conducted on 200 randomly sampled APIs (100 for TensorFlow and 100 for PyTorch).
Inspired by an existing work~\cite{tensorscope}, we run our test input generation module in a fixed time budget (\ie, five hours) based on path constraints extracted from both \toolname{} and \toolnamenonllm{}.
Figure~\ref{fig:ablation_coverage_trend} shows the overall branch coverage achieved by \toolname{} and \toolnamenonllm{} for both TensorFlow and PyTorch.
After running our test input generation for five hours, inputs generated following path constraints from \toolname{} outperforms those from \toolnamenonllm{} by covering 6.10\% more branches (15,889 v.s. 14,975, considering branches in both TensorFlow and PyTorch).
It is worth noting that a substantial portion of DL library branches (\eg, those handling API initialization) are covered at the beginning of testing (see Figure~\ref{fig:ablation_coverage_trend}).
This observation aligns with experiment results from prior API-level testing works~\cite{acetest,tensorscope}. 
Beyond this initial coverage (approximately 12,800 branches), \toolnamenonllm{} covers only 2,175 additional branches during the remaining five-hour testing.
In contrast, \toolname{} achieves a significantly higher coverage improvement, covering 3,089 new branches.
This result demonstrates the value of inferred input constraints in test input generation.
Despite not all constraints being accurate, the LLM-based input constraint inference could infer a substantial number of correct input constraints, which could effectively narrow down the search space to improve the effectiveness of test input generation.

\begin{mdframed}[style=MyFrame]
\textbf{Summary of RQ2}:
Overall, \toolname{} outperforms the state-of-the-art constraint extraction approach (DocTer~\cite{docter}) by extracting 45.07\% more property constraints, especially in extracting constraints for shape, value, and structure properties.
By comparing \toolname{} with our baseline \toolnamenonllm{} (extracting path constraints without using LLM), we notice that our proposed LLM-based input constraint inference can significantly enhance \toolname{} in the number of extracted path constraints and the number of input constraints per path.
Despite not all of our inferred constraints being correct, our ablation study on 200 randomly sampled APIs demonstrates that these inferred constraints can effectively guide test input generation to achieve a higher branch coverage, compared to path constraints not including these inferred constraints.
\end{mdframed}

\subsection{RQ3: Comparison on Bug Detection and Branch Coverage on Randomly Sampled APIs}

\subsubsection{Metrics}
We use the number of detected bugs and branch coverage as the evaluation metrics.

\begin{itemize}[leftmargin=*,topsep=0pt]
    \item \textbf{Number of Detected Bugs}. During the bug detection, we manually analyze all bugs reported by each tool and filter out false positives. 
    For each tool, we reported the number of unique bugs detected.
    \item \textbf{Branch coverage in sampled APIs}. Branch coverage has been widely adopted in recent DL library testing techniques~\cite{acetest,comet} to reveal the path exploration ability of test cases~\cite{acetest}. 
    We follow a recent work (COMET~\cite{comet}) to use the coverage.py~\cite{coverage_py} and lcov~\cite{lcov} to collect branch coverage for Python and C/C++ code, respectively.
    When measuring the branch coverage, we only include branches covered during the execution of APIs under test. 
    To ensure a fair comparison, additional branches introduced by irrelevant APIs before executing APIs under test are excluded.
\end{itemize}

\subsubsection{Setup}
We compare \toolname{} with five state-of-the-art DL library testing tools: ACETest~\cite{acetest}, TitanFuzz~\cite{titanfuzz}, DocTer~\cite{docter}, DeepREL~\cite{deeprel}, and FreeFuzz~\cite{freelunch}.
To conduct this comparison, we randomly sample 100 TensorFlow APIs and 100 PyTorch APIs from the total 689 TensorFlow APIs and 712 PyTorch APIs collected in RQ1.
The versions of DL libraries are TensorFlow v2.10.0 and PyTorch v2.1.0.
Following the existing setting~\cite{tensorscope}, we execute each tool to generate test inputs and perform bug detection on these sampled APIs under the same time budget (five hours) and testing environment.

\begin{figure}[htbp]
\centering
\includegraphics[width=0.9\linewidth]{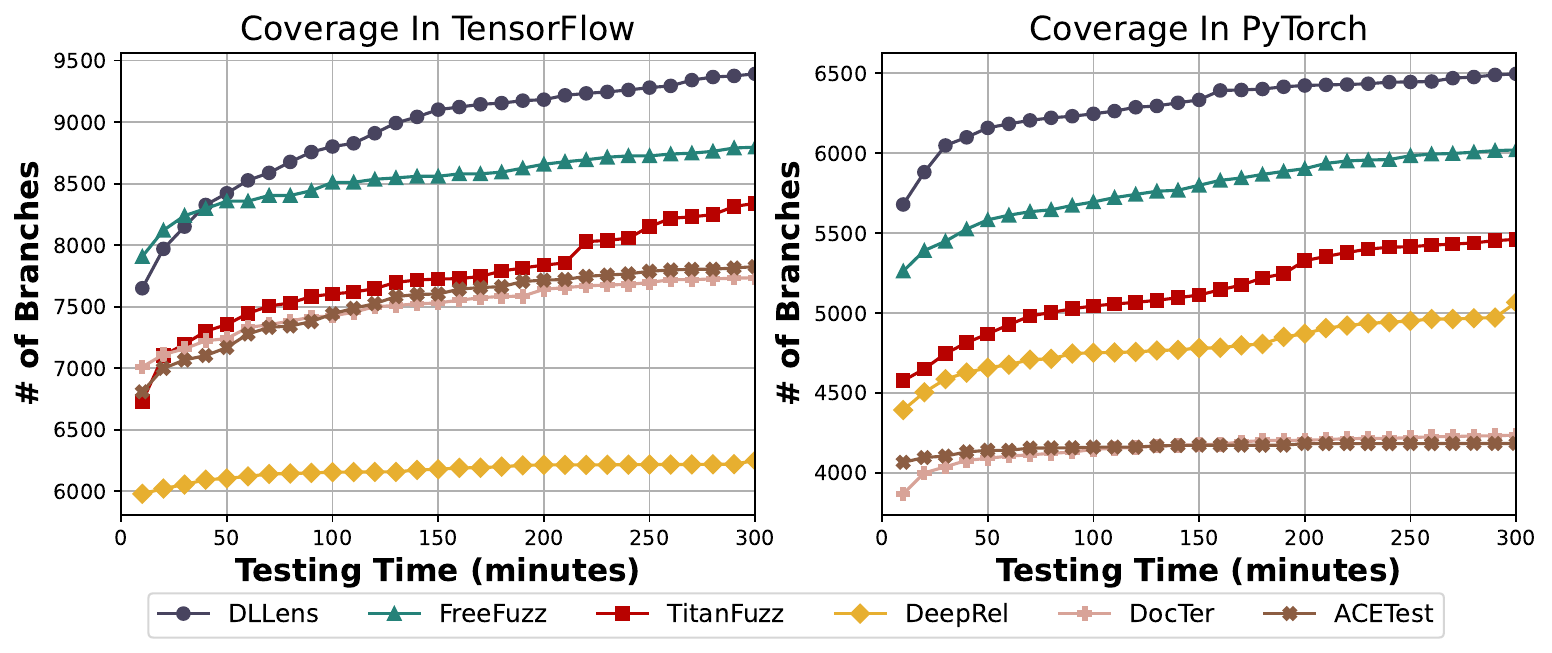}
\caption{Comparison of code coverage between \toolname{} and baselines.}\label{fig:rq4_coverage_trend}
\Description{RQ3 Code Coverage}
\end{figure}

\subsubsection{Results}
Figure~\ref{fig:rq4_coverage_trend} compares the coverage trend of \toolname{} with existing works.
\toolname{} achieves the highest branch coverage in most of the time, and the number of covered branches increases steadily along execution time.
After running for five hours, \toolname{} outperforms the best baseline by covering 7.23\% (15,889 v.s. 14,817) more branches, including 6.78\% more branches (9393 v.s. 8797) in TensorFlow and 7.91\% more branches (6496 v.s. 6020) in PyTorch.
Notably, \toolname{} can already cover 7650 and 5679 PyTorch branches for TensorFlow and PyTorch APIs at the beginning of testing. 
This result is comparable to the best baseline (FreeFuzz), which relies on a large number of test inputs from open source (\eg, documentation, unit test cases) for testing.
This result highlights the usefulness of path constraints extracted by \toolname{}, which could effectively narrow down the search space so many execution paths can be reached through quick searching.
As the testing time increases, \toolname{} outperforms all existing approaches by effectively exploring more branches.
In contrast, the number of branches covered by the best baseline (FreeFuzz), which generates inputs using random mutation, saturates quickly after 50 minutes of testing.
This demonstrates the effectiveness of \toolname{} in continuously generating diverse test inputs to explore library branches.

Additionally, we notice that ACETest~\cite{acetest} and DocTer~\cite{docter} only cover a limited number of PyTorch branches (\textasciitilde4200).
One possible reason is the limited number of PyTorch APIs that they support.
Both ACETest and DocTer rely on input constraints for test input generation.
If constraints are not provided, these tools may fail to generate test inputs for the concerned APIs.
In our experiment, among the 100 randomly sampled PyTorch APIs, these tools only include constraints for 49 and 34 PyTorch APIs, respectively.
As a result, ACETest fails to test 51 out of these 100 PyTorch APIs and DocTer fails to test 66 out of them.

\begin{table}[htbp]
\caption{Comparison of bug detection between \toolname{} and baselines.}\label{tab:rq4_bug_detection}
\resizebox{0.5\linewidth}{!}{
\begin{tabular}{l|r|r|r|r@{}}
\toprule
Tool       & Inc.Rej. & Incorrect & Crash  & Total Bugs   \\ \midrule
DocTer*     & 0        & 0        & 0       & 0            \\ \midrule
DeepREL    & 0        & 0        & 2       & 2            \\ \midrule
TitanFuzz  & 0        & 0        & 4       & 4            \\ \midrule
ACETest    & 0        & 0        & 4       & 4             \\ \midrule
FreeFuzz   & 5        & 3        & 0       & 8            \\ \midrule
\toolname{}& 5        & 7        & 3       & 15             \\ \bottomrule
\end{tabular}
}
\end{table}

Table~\ref{tab:rq4_bug_detection} compares the bug detection performance of \toolname{} with existing tools.
After five hours running on 200 APIs, \toolname{} outperforms all five state-of-the-art tools. 
It detects a total of 15 bugs, surpassing the best baseline (FreeFuzz) by finding 7 more bugs.
It is noticed that most baselines (except FreeFuzz) detect less than five bugs.
One possible reason is that these tools mainly focus on detecting crash bugs, while their test oracles may be limited when detecting functional bugs, including incorrect result bugs and incorrectly rejected bugs.
In contrast, \toolname{} have detected 5 incorrectly-rejected bugs and 7 incorrect result bugs.
These bugs are detected by checking output inconsistencies between the API under test and its counterpart, demonstrating the usefulness of counterparts synthesized by \toolname{}.
Besides these 12 bugs, \toolname{} has also detected 3 crash bugs.
Crash bugs are usually triggered by carefully crafted test inputs that can pass the input validity check to reach the core logic (as exemplified in Listing~\ref{lst:rq4_crash_bug}).
Therefore, detecting these crash bugs demonstrates the usefulness of path constraints extracted by \toolname{} in guiding test input generation to reach the core logic.

We observe that DocTer detects no bug inside the sampled 200 APIs.
One possible reason is that DocTer relies on pre-defined rules to extract constraints from documentation, which may not be applicable to all APIs.
After careful investigation, we discover that constraints released by DocTer~\cite{docter_repo} only cover 83 out of 200 sampled APIs in this RQ, including 49 TensorFlow APIs and 34 PyTorch APIs.
For APIs lacking constraints, DocTer will not generate test inputs, leaving 117 sampled APIs not tested.

\begin{mdframed}[style=MyFrame]
\textbf{Summary of RQ3}:
\toolname{} outperforms all existing tools in terms of the number of detected bugs and the number of covered branches on 200 randomly sampled APIs.
After running five hours for each library, \toolname{} covers 6.78\% more branches in the sampled TensorFlow APIs and 7.91\% more branches in the sampled PyTorch APIs than the best-performed baseline.
Regarding the number of detected bugs, \toolname{} detects 15 bugs, surpassing the best-performed baseline by 7 bugs.
Notably, most bugs (12) detected by \toolname{} are exposed via checking output inconsistencies between the API under test and its counterpart, demonstrating the usefulness of counterparts synthesized by \toolname{}.
\end{mdframed}

\subsection{RQ4: Summary of Detected Bugs}
Table~\ref{tab:evaluation-bug} summarizes the statistics of bugs detected by \toolname{}.
In total, \toolname{} detected \reportedbugs{} bugs, with \confirmedbugs{} confirmed, including \newbugs{} confirmed as previously unknown bugs.
The remaining \rejectedbugs{} bugs were rejected by the DL library developers.
All our detected bugs are triaged by developers (\eg, no bugs are pending confirmation).
Among the \newbugs{} previously unknown bugs, \fixedbugs{} of them have already been fixed in the latest version of TensorFlow and PyTorch.
We further present the number of confirmed bugs for each bug type (see Table~\ref{tab:confirmed_bug_distribution}).
\begin{table}[htbp]
\caption{Summary of detected bugs.}\label{tab:evaluation-bug}
\newcommand{\twocol}[1]{\multicolumn{2}{c}{#1}}
\newcommand{\tworow}[1]{\multirow{2}{*}{#1}}
\newcommand{\fourcol}[1]{\multicolumn{4}{c}{#1}}
\centering
\renewcommand{\arraystretch}{1.0}
\resizebox{0.55\linewidth}{!}{
\begin{threeparttable}
\begin{tabular}{@{}l|r|rrrr@{}}
\toprule
\tworow{Library Name} &  \tworow{Total}& \twocol{Confirmed}  & \tworow{Rejected}  & \tworow{Pending}  \\ \cmidrule{3-4}
             &                &  Known  & Unknown   &                    &          \\ \midrule
TensorFlow   & 46 (7)         &  6      & 31 (7)    & 9                  & 0        \\ \midrule
PyTorch      & 25 (3)         &  7      & 15 (3)    & 3                  & 0        \\ \midrule
Subtotal     & \reportedbugs{} (\fixedbugs{})        & \knownbugs{}      & \newbugs{} (\fixedbugs{})   & \rejectedbugs{}                  & 0        \\ \bottomrule
\end{tabular}
\end{threeparttable}
}
\end{table}

\begin{table}[htbp]
\caption{Symptoms of confirmed bugs.}\label{tab:confirmed_bug_distribution}
\newcommand{\twocol}[1]{\multicolumn{2}{c|}{#1}}
\newcommand{\tworow}[1]{\multirow{2}{*}{#1}}
\newcommand{\fourcol}[1]{\multicolumn{4}{c||}{#1}}
\centering
\renewcommand{\arraystretch}{1.0}
\resizebox{0.9\linewidth}{!}{
\begin{threeparttable}
\begin{tabular}{@{}l|r|rrrrrr@{}}
\toprule
\tworow{Library Name} & \tworow{Total Confirmed}                         & \twocol{Inc. Rej.} & \twocol{Incorrect} & \multicolumn{2}{c}{Crash}   \\ \cmidrule{3-8}
                      &                                                  & Known & Unknown    & Known  &  Unknown  & Known  & Unknown \\ \midrule
TensorFlow   & 37 (7)                                           & 1     & 8 (1)     & 2      & 18 (6)    & 3      & 5 (0)   \\ \midrule
PyTorch      & 22 (3)                                           & 0     & 0         & 3      & 8 (2)     & 4      & 7 (1)   \\ \midrule
Subtotal     & \confirmedbugs{} (\fixedbugs{})                  & 1     & 8 (1)    & 5      & 26 (8)    & 7      & 12 (1)  \\ \bottomrule
\end{tabular}
\begin{tablenotes}
\item [1] In Table~\ref{tab:evaluation-bug} and Table~\ref{tab:confirmed_bug_distribution}, the number of fixed bugs are in parentheses.
\end{tablenotes}
\end{threeparttable}
}
\end{table}

Among all \confirmedbugs{} confirmed bugs, 52.54\% (31 out of \confirmedbugs{}) bugs detected by \toolname{} are incorrect result bugs (`Incorrect' in Table~\ref{tab:confirmed_bug_distribution}), 15.25\% (9 out of \confirmedbugs{}) are incorrectly-rejected bugs (`Inc.Rej.' in Table~\ref{tab:confirmed_bug_distribution}).
The detection of both incorrect result bugs and incorrectly-rejected bugs requires the test oracle, which can be effectively addressed by our synthesized counterparts.
We notice that \toolname{} has also detected many crash bugs (`Crash' in Table~\ref{tab:evaluation-bug}), which take up to 19 out of \confirmedbugs{}.
Although a counterpart is not necessary to detect these bugs, triggering these bugs often requires carefully-crafted test inputs to pass the input validity check and reach the core function logic of the DL library APIs~\cite{acetest}.
Hence, the detection of these crashes demonstrates the usefulness of path constraints extracted by \toolname{} in guiding test input generation to reach the core logic.
Next, we present example bugs for each symptom, as well as one rejected bug.

\begin{listing}[htbp]
\begin{minted}[
    baselinestretch=1.0,
    fontsize=\scriptsize,
    xleftmargin=0.5ex,
    bgcolor=bg,
    breaklines=true,
    escapeinside=||,
]{text}
|\underline{\textbf{Bug Triggering Input}}|:
x = tf.constant([10,9], dtype='uint32')
|\underline{\textbf{Buggy API}}|:
out = tf.math.is_non_decreasing(x)
|\underline{\textbf{Actual Result (Expected Result)}}|:
|\textbf{\textcolor{bgRed}{True}}| (|\textbf{\textcolor{bgGreen}{False}}|)
|\underline{\textbf{Synthesized Counterpart}}|:
def pytorch_call(x):
    return torch.all(torch.eq(x, torch.sort(x)[0]))
|\underline{\textbf{Output of Counterpart}}|:
|\textbf{\textcolor{bgGreen}{False}}|
\end{minted}
\caption{An incorrect result bug detected by \toolname{}.}
\label{lst:rq4_incorrect_result_bug}
\end{listing}

\begin{listing}[htbp]
\begin{minted}[
    baselinestretch=1.0,
    fontsize=\scriptsize,
    xleftmargin=0.5ex,
    bgcolor=bg,
    breaklines=true,
    escapeinside=||,
]{text}
|\underline{\textbf{Bug Triggering Input}}|:
x = tf.constant(np.random.randint(-50, 50, ()), dtype='float16')
y = tf.constant(np.random.randint(-50, 50, ()), dtype='float16')
|\underline{\textbf{Buggy API}}|:
tf.raw_ops.ApproximateEqual(x=x, y=y)
|\underline{\textbf{Error Message}}|:
|\textbf{\textcolor{bgRed}{NotFoundError: Could not find device for node: {{node ApproximateEqual}} = ApproximateEqual[T=DT\_HALF, tolerance=1e-05]}}|
|\underline{\textbf{Synthesized Counterpart}}|:
def pytorch_call(x, y, tolerance=1e-05):
    return torch.isclose(x, y, atol=tolerance)
|\underline{\textbf{Execution Status of Counterpart}}|:
|\textbf{\textcolor{bgGreen}{No Exception}}|
\end{minted}
\caption{An incorrectly-rejected bug detected by \toolname{}.}
\label{lst:rq4_incorrectly_rejected_bug}
\end{listing}

\begin{listing}[htbp]
\begin{minted}[
    baselinestretch=1.0,
    fontsize=\scriptsize,
    xleftmargin=0.5ex,
    bgcolor=bg,
    breaklines=true,
    escapeinside=||,
]{text}
|\underline{\textbf{Bug Triggering Input}}|:
input = tf.constant([248, 225])
indices = tf.constant([[[1]]])
updates = tf.constant([16])
|\underline{\textbf{Buggy API}}|:
tf.tensor_scatter_nd_update(input, indices, updates)
|\underline{\textbf{Symptom}}|:
|\textbf{\textcolor{bgRed}{Program Abort}}|
|\underline{Two Validation Checks Before Reaching The Buggy Code}|:
OP_REQUIRES(.., |\textbf{indices.shape().dims() >= 1}|,..);
OP_REQUIRES(.., |\textbf{updates.shape().dims() >= 1}|,..);
for (int i = 0; i < outer_dims; ++i) { ... // buggy code
\end{minted}
\caption{A crash bug detected by \toolname{}.}
\label{lst:rq4_crash_bug}
\end{listing}

\begin{listing}[htbp]
\begin{minted}[
    baselinestretch=1.0,
    fontsize=\scriptsize,
    xleftmargin=0.5ex,
    bgcolor=bg,
    breaklines=true,
    escapeinside=||,
]{text}
|\underline{\textbf{Test Input}}|:
x = tf.constant([np.inf+0.j], dtype='complex64')
|\underline{\textbf{Buggy API}}|:
tf.keras.activations.sigmoid(x)
|\underline{\textbf{Counterpart}}|:
def pytorch_call(x):
    return torch.sigmoid(x)
|\underline{\textbf{API's Output v.s. Counterpart's Output}}|:
|\textbf{nan+nanj}| v.s. |\textbf{1.+0.j}|
|\underline{\textbf{Developer's Comment}}|:
The tf.math.sigmoid function outputs NaN when it receives an infinity complex tensor as input. 
This behavior is expected and aligns with the mathematical properties of the sigmoid function.
\end{minted}
\caption{A bug rejected by a DL library developer.}
\label{lst:rq4_rejected_bug}
\end{listing}

Listing~\ref{lst:rq4_incorrect_result_bug} shows an example of an incorrect result bug detected by \toolname{}.
The buggy API {\mycode{}tf.math.is\_non\_decreasing} is designed to check if a given tensor follows the non-decreasing order (\ie, for tensor $[x[0],\ldots], x[i]\leq x[i+1]$), which is a common DL functionality~\cite{python_list_decreasing}.
However, it incorrectly outputs {\mycode{}True} when receiving a {\mycode{}uint32} tensor with value to be $[10,9]$, which violates the non-decreasing order~\cite{tf_is_non_decreasing}.
This bug is possibly caused by an internal overflow that occurs during the computation of this API\@.
After reporting this bug, we received a comment from the TensorFlow team: ``\textit{Thanks for reporting this. $\dots$ We will check and submit a fix for this. Thank you!}'' 
This bug is promptly fixed by the TensorFlow team.
Existing work cannot detect this bug since counterparts identified by them produce the same incorrect result as the buggy API\@.
It is noteworthy that the counterpart synthesized by \toolname{} can correctly compute the result, which is {\mycode{}False}.

An example of an incorrectly-rejected bug is demonstrated in Listing~\ref{lst:rq4_incorrectly_rejected_bug}.
The buggy throws a {\mycode{}NotFoundError} when receiving a {\mycode{}float16} input, which violates its documentation~\cite{tf_approximate_equal}.
Although the fault-triggering input seems easy to generate, detecting this bug requires the test oracle to check if this Exception is thrown correctly.
This bug is exposed since the synthesized counterpart can successfully perform the computation without raising an exception.
After reporting this bug, the TensorFlow team has initiated a pull request to fix it.

Listing~\ref{lst:rq4_crash_bug} is an example of the crash bug when the buggy API raises a crash when the {\mycode{}ndims} of {\mycode{}indices} is 3.
Reaching the buggy code requires passing two sanity checks which assert the {\mycode{}indices.ndims} and {\mycode{}updates.ndims} to be no less than 1~\cite{tf_tensor_scatter_nd_update}.
\toolname{} can detect this bug since it can successfully extract the constraints required by these two sanity checks and guide test input generation to reach the buggy code, increasing the probability of triggering this bug.
A pull request has been initiated by the TensorFlow team to fix this bug.

Lastly, we present a bug rejected by the DL library developer (Listing~\ref{lst:rq4_rejected_bug}).
When receiving a {\mycode{}complex64} tensor containing an infinity value, {\mycode{}tf.keras.activation} outputs NaN result, which is inconsistent with the counterpart's output~\cite{tf_sigmoid_nan_rejected}.
We report this bug because outputting NaN seems to violate the computation of sigmoid equation $\frac{1}{(1+e^{-x})}$ when $x$ is infinity.
The possible reason for this bug to be rejected is that TensorFlow contains internal handling of extreme values like {\mycode{}np.inf} and considers raising NaN to be a reasonable output when some API's input is {\mycode{}np.inf}.

\begin{mdframed}[style=MyFrame]
\textbf{Summary of RQ4}:
In conclusion, \toolname{} detected \reportedbugs{} bugs, with \confirmedbugs{} confirmed, including \newbugs{} confirmed as previously unknown bugs.
More importantly, we find that 40 out of \confirmedbugs{} confirmed bugs are detected via checking the output consistency between the API under test and its counterpart.
This finding further supports the usefulness of synthesized counterparts in detecting functional bugs in DL libraries.
\end{mdframed}

\section{Discussions}\label{sec:discussions}
\subsection{Synthesizing Counterparts Using Other Sources}
Although \toolname{} uses a different DL library as the source in synthesizing counterparts for DL library APIs, 
the methodology could be extended to non-DL sources like NumPy that can also be used to realize DL API functionalities.
For example, NumPy’s array operations (\eg, matrix multiplications) are commonly used in DL algorithms.
It is interesting to understand if considering more sources (\eg, NumPy) could contribute to more valid counterparts.
To empirically validate this feasibility, we extend \toolname{} to synthesize TensorFlow API counterparts using NumPy and compared results against its original PyTorch-based synthesis.

\begin{figure}[t!]
    \centering
    \includegraphics[width=0.5\textwidth]{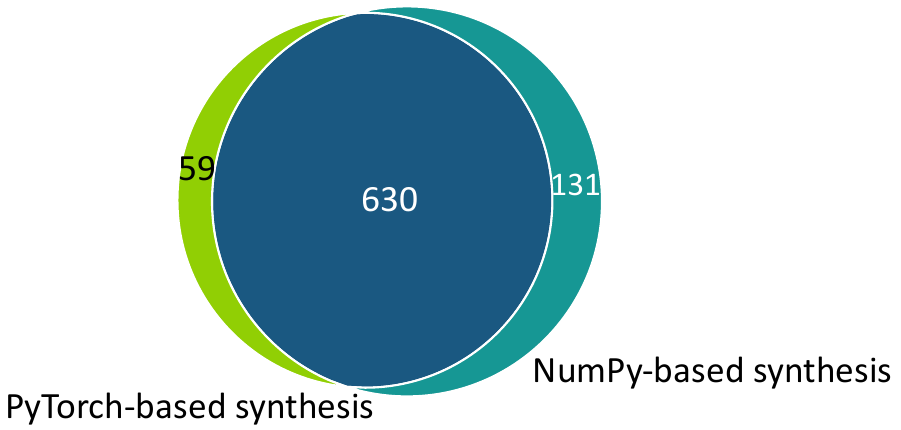}
    \caption{The number of APIs that can be successfully synthesized using NumPy APIs and PyTorch APIs.}
    \label{fig:discussion_torch_np_counterpart}
\end{figure}

Figure~\ref{fig:discussion_torch_np_counterpart} compares the number of TensorFlow APIs whose counterparts can be successfully synthesized by \toolname{}, using different sources (\ie, NumPy or PyTorch).
After switching counterpart synthesis sources from PyTorch to NumPy, we notice that \toolname{} performs similarly well: \toolname{} can successfully synthesize counterparts for 761 APIs using NumPy-based synthesis.
Moreover, both NumPy and PyTorch sources are complementary to each other: NumPy-based synthesis uniquely finds counterparts for 131 APIs, whose counterparts were not found by PyTorch-based synthesis; and PyTorch-based synthesis uniquely finds counterparts for 59 APIs.
This demonstrates the extensibility of our counterpart synthesis approach, different sources could collectively contribute to more valid counterparts, as different libraries may provide distinct functionalities.
While our current implementation prioritizes DL libraries as synthesis sources for practical testing efficiency, this framework can theoretically incorporate other sources to synthesize counterparts.

\subsection{False Positives in Bug Detection}
We summarize three primary types of false positives: Firstly, some false positives are introduced by partially valid counterparts synthesized by \toolname{}.
These partially valid counterparts may perform inconsistent functionality for certain inputs (as is exemplified in Listing~\ref{lst:examplecounterpart_partially_valid} of RQ1).
As a result, the output inconsistencies \textit{w.r.t} these inputs between APIs under test and their counterparts may be false positives.
To understand the influence of these false positives, we manually analyze all inconsistencies reported by \toolname{} in RQ3, which includes 200 APIs and their counterparts.
According to our analysis, only 24 false positive inconsistencies reported by \toolname{} are caused by partially valid counterparts, which cost \textasciitilde2 hours to filter out.
Considering the effectiveness of \toolname{} in detecting real bugs, we believe such overhead introduced by these partially valid counterparts is affordable.

Secondly, false positives could be introduced by inconsistent computation precision between APIs and their counterparts.
For instance, given a same test input on which an API under test performs {\mycode{}float32} precision computation, this API's counterpart may perform {\mycode{}float16} precision computation.
Such computation precision difference could lead to false positive inconsistencies.
Regarding counterparts introducing this type of false positive, we find a common pattern that they often perform redundant type castings, which may diverge their computation result from that of the APIs under test.
In total, 8 false positives are caused by inconsistent computation precision, which take \textasciitilde20 minutes to filter out.

Another type of false positive is introduced by inconsistent behavior between libraries when handling special values such as {\mycode{}NaN}~\cite{example}.
For inconsistencies introduced by these special values, we manually check the documentation of the API under test with together with all APIs in its counterpart and filter out false positives.
In total, 14 false positives are caused by these special values, which take \textasciitilde1 hour to filter out.

\subsection{Future Directions to Reduce False Positives}
According to our analysis, the false positives in bug detection are mainly caused by implementation differences between the API under test and its counterpart.
This observation is consistent with the observation from existing DL library differential testing works~\cite{deeprel,audee,gandalf}.
Although entirely eliminating these false positives automatically is challenging, we summarize possible research directions to reduce them, hoping to inspire future DL library differential testing works.

\textbf{Research direction 1: Improving the quality of counterparts}. As discussed in Section 5.2 and pointed out by existing works~\cite{audee,gandalf}, counterparts used for differential testing may not always be valid (see Listing~\ref{lst:examplecounterpart_partially_valid} as an example).
Reducing the number of partially valid counterparts or enlarging the valid input range of synthesized counterparts could help reduce false positives.

\textbf{Research direction 2: Understanding the valid input range of counterparts}.
While generating perfect counterparts is challenging, we observe that the counterpart could still be useful if the test input is within its valid input range.
Inspired by this observation, another promising direction is to understand the valid input range of counterparts.
By doing so, research works could guide test input generation to generate input within this range, thus reducing false positives.

\subsection{Threat to Validity}
Our experiment results are subject to three threats.
The first threat is that the LLM's hallucination issue may affect the quality of our synthesized counterparts, thus introducing false positive inconsistencies during bug detection.
To mitigate the influence of this issue, we use a diverse set of validation inputs to validate LLM's synthesized counterparts and only output those counterparts passing these validation inputs.
We also conduct a manual analysis on our synthesized counterparts to better understand the quality of our counterparts.
Our analysis result shows that most of our synthesized counterparts are correct and only a few may lead to false positive inconsistencies on certain inputs during the bug detection.
During bug detection, we further manually filter out these false positive inconsistencies.

The second threat concerns the time budget configuration of five hours for each tool on each library in RQ3. 
The amount of available system resources for the running jobs may fluctuate during the experimentation.
To alleviate the influence caused by system computing performance, we ensure the server's computation resources are sufficient when running each tool.
Besides, we observe that the number of covered branches is nearly saturated after five hours, which implies that the fuzzing budget is sufficient for each tool.

The third threat is introduced by our branch coverage collection measurement.
To include only the branches covered during the execution of the APIs under test, we use a Python standard Pickle package to temporarily store the APIs' test inputs incurred when running each tool.
We collect the branch coverage by loading these test inputs and feeding them to the concerned APIs. In this way, no external DL library APIs will be called during the branch coverage collection.
We also exclude model objects as test inputs because they will trigger many model-related branches (\eg, model loading branches) that are not related to the APIs under test.
Since existing works do not explain the procedures of collecting the branches/lines covered in each API's implementation, our branch coverage collection process may differ from theirs, thus may not fully capture their performance. 
To ensure the fairness of our comparison, we conduct the same branch coverage collection process for all tools so each tool's branch coverage is measured under the same setting.

The last threat is introduced by our counterpart validation. 
Our counterpart validation may mistakenly reject some valid counterparts if validation inputs trigger real inconsistency bugs.
This challenge is inherent to our approach and prior work~\cite{deeprel}, which applies a similar counterpart validation design.
While balancing false positives (\ie, accepting invalid counterparts) and false negatives (\ie, rejecting valid counterparts) is challenging, we align our design with DeepREL~\cite{deeprel} to prioritize validation precision for the reliability of synthesized counterparts.
Although this design choice could miss some valid counterparts, our experiment results demonstrate that this design remains highly effective in practice: \toolname{} detected 71 bugs in TensorFlow and PyTorch, suggesting that these false negatives do not significantly hinder the real bug detection capability of \toolname.

\section{Related Works}
\subsection{DL Library Testing}
\label{related:dl_library_testing}
Existing testing approaches for DL libraries focus on either the model level or API level. 
The former~\cite{cradle,audee,lemon,graphfuzz,muffin,comet} takes various DL models as test inputs and tries to exercise specific modules (\eg, model construction, model training, and inference) in DL libraries.
For instance, CRADLE~\cite{cradle} use existing DL models such as ResNet50~\cite{resnet} as test inputs and compare the model's inference outputs on multiple DL libraries.
LEMON~\cite{lemon}, Audee~\cite{audee}, GraphFuzz~\cite{graphfuzz}, Muffin~\cite{muffin}, and COMET~\cite{comet} further proposed a set of mutation operators to generate diverse DL models for testing.
However, a recent study has revealed that these model-level testing approaches are ineffective in test adequacy~\cite{freelunch}. 
One potential explanation is that these approaches can operate on and manipulate only layer APIs within DL libraries, leaving the majority of library APIs unexplored~\cite{freelunch,titanfuzz}.

API-level testing, which improves the coverage of DL library APIs by directly executing them, provides an alternative approach to testing DL libraries.
Several approaches~\cite{freelunch,docter,acetest} have been introduced to generate input arguments (\ie, test inputs) to test these APIs.
FreeFuzz~\cite{freelunch} collects API input arguments from open source code (\eg, unit test suite written by developers), and further randomly mutates these inputs for testing.
DocTer~\cite{docter} and ACETest~\cite{acetest} extracts API's input constraints from its documentation or input validation code and then uses these constraints to generate inputs for testing.
More recently, TitanFuzz~\cite{titanfuzz} and FuzzGPT~\cite{fuzzgpt} leverage large language models (LLMs) like ChatGPT for generating input arguments for DL library APIs. 
Compared with existing API-level testing approaches, \toolname{} stands by effectively synthesizing counterparts for DL library APIs so differential testing can be applied across different libraries.
\toolname{} is also able to generate diverse test inputs to explore execution paths inside DL library APIs.

\subsection{Differential Testing For DL Libraries Testing}\label{related:differential_testing}
Existing approaches, as mentioned in \cref{subsec:limitation_counterparts}, mostly adopt differential testing to check output consistency between DL library APIs and their counterparts.
There are two widely used paradigms to find counterparts.
The first paradigm~\cite{cradle,tensorscope} leverages model conversion tools such as TF2ONNX~\cite{tf2onnx} to compare outputs across \textit{multiple} DL libraries.
For instance, given a DL library API, TensorScope~\cite{tensorscope}, parsed the conversion rules in TF2ONNX~\cite{tf2onnx} to extract the counterpart of this API, which is implemented in another DL library.
The second paradigm applies differential testing by finding counterparts within a \textit{single} DL libraries.
For instance, several works~\cite{freelunch,lambdafuzz,eagle,titanfuzz,fuzzgpt} detect functional bugs via checking if an API can have consistent outputs under different computation modes, such as backends (\eg, CPU and GPU), and execution modes (\eg, different gradient calculation modes).
DeepRel~\cite{deeprel} mined multiple groups of similar APIs within the same DL libraries based on API signature and document similarity.

However, the first paradigm can only support counterparts for a limited number of DL library APIs. 
For example, TF2ONNX~\cite{tf2onnx} only supports counterparts for only 304 TensorFlow APIs.
For the second paradigm, counterparts identified within the same DL library may yield identical computation results with their APIs, making checking result inconsistency for bug detection ineffective.
For instance, {\mycode{}tf.argmax} and {\mycode{}tf.math.argmax} found by DeepReal are indeed aliases, suggesting that they are likely to have the same outputs.
This paper follows the first paradigm to target synthesizing counterparts for each DL library API from a another DL libraries.
Compared with existing approaches in the first paradigm, \toolname{} does not rely on existing model conversion libraries, and it can synthesize counterparts for more library APIs.

\subsection{Constraint Extraction for DL Library Testing}
Some approaches have proposed constraint extraction strategies~\cite{docter,acetest,tensorscope} to guide valid/invalid input generation.
DocTer~\cite{docter} collected constraints from DL library documentation and generated inputs that follow these constraints (denoted as valid inputs) and inputs violating these constraints.
TensorScope~\cite{tensorscope} parsed the input validation check statements to generate valid inputs.
However, both TensorScope and DocTer~\cite{docter} do not provide the constraints required for the different execution paths prescribed by the implementation of a DL library API\@.

ACETest~\cite{acetest} proposed a static analysis tool that extracted path constraints from the validation code inside DL library APIs, so input following these path constraints can pass the validation code to reach the core logic inside these APIs.
However, due to the large code size of DL library APIs and the scalability issue of static analysis, ACETest~\cite{acetest} could not perform static analysis on some functions (\ie, functions from external libraries or modules), resulting in many incomplete path constraints.
Compared with ACETest, \toolname{} leverages an LLM to enhance static analysis by analyzing those external functions, thus it can extract more complete path constraints.

WhiteFox~\cite{whitefox} is the latest work that extracts constraints from DL library source code via LLMs.
While WhiteFox similarly employs LLMs to analyze DL library code for constraint extraction, its focus diverges from \toolname{} in objective and methodology. 
Specifically, WhiteFox targets triggering model optimization bugs inside DL libraries, using LLMs to both extract high-level constraints (\eg, API sequences that could trigger target model optimizations) and generate concrete models triggering these optimizations.
In contrast, \toolname{} focuses on API-specific bugs, it uses an LLM to extract path constraints (\eg, tensor shape constraints) from API implementations, then leverages Z3 to solve them for API input generation.
Therefore, the scopes of WhiteFox and \toolname{} are complementary: WhiteFox broadly exercises optimization logic through generated models, while \toolname{} explores paths inside APIs via diverse test input generations.
In addition, compared with WhiteFox, \toolname{} requires more precise constraints for Z3 solver-based test input generation, while WhiteFox's LLM-driven model generation is more flexible but requires more generation time (due to the LLM's inference overhead).

\section{Conclusion}
In this paper, we propose a novel technique named \toolname{} to test DL libraries
\toolname{} facilitates differential testing by using a novel counterpart synthesis method and generates diverse test inputs to explore execution paths via an effective path constraint extraction method.
\toolname{} can synthesize counterparts for 1.84 times as many APIs as those found by state-of-the-art techniques.
Benefiting from the synthesized counterparts and effective path constraint extraction, \toolname{} outperforms state-of-the-art approaches in terms of bug detection and branch coverage in a set of 200 randomly sampled APIs.
In total, \toolname{} detects \reportedbugs{} including \confirmedbugs{} bugs confirmed.
\newbugs{} of these confirmed bugs are previously unknown bugs.
So far, \fixedbugs{} of our reported previously unknown bugs have been fixed in the latest version of TensorFlow and PyTorch.

\section{Data Availability}
We release the implementation and associated experiment data in a public GitHub repository~\cite{DLLens}.
\section*{Acknowledgements}
We would like to thank the anonymous reviewers for their comments and suggestions. 
We would also like to thank DL library developers for analyzing our reported issues. 
This work was supported by the Hong Kong SAR Research Grant Council/General Research Fund (Grant No. 16205722).

\bibliographystyle{ACM-Reference-Format}
\bibliography{reference}
\end{document}